\documentclass[review]{elsarticle}
\usepackage[acronym,toc]{glossaries}
\newacronym{MIT}{MIT}{the Massachusettes Institute of Technology}
\newacronym{UW}{UW}{University of Wisconsin}
\newacronym{US}{US}{United States}
\newacronym{IAEA}{IAEA}{International Atomic Energy Agency}
\newacronym{SNF}{SNF}{spent nuclear fuel}
\newacronym{HLW}{HLW}{high level waste}
\newacronym{FEHM}{FEHM}{Finite Element Heat and Mass Transfer}
\newacronym{DOE}{DOE}{Department of Energy}
\newacronym{GENIUSv1}{GENIUS}{Global Evaluation of Nuclear Infrastructure 
Utilization Scenarios, Version 1}
\newacronym{GENIUSv2}{GENIUS}{Global Evaluation of Nuclear Infrastructure Utilization Scenarios, Version 2}
\newacronym{CNERG}{CNERG}{Computational Nuclear Engineering Research Group}
\newacronym{GDSM}{GDSM}{Generic Disposal System Model}
\newacronym{GDSE}{GDSE}{Generic Disposal Sytem Environment}
\newacronym{GPAM}{GPAM}{Generic Performance Asessment Model}
\newacronym{FEPs}{FEPs}{Features, Events, and Processes}
\newacronym{EBS}{EBS}{Engineered Barrier System}
\newacronym{EDZ}{EDZ}{Excavation Disturbed Zone}
\newacronym{YMR}{YMR}{Yucca Mountain Repository Site}
\newacronym{EPA}{EPA}{Environmental Protection Agency}
\newacronym{PEI}{PEI}{Peak Environmental Impact}
\newacronym{VISION}{VISION}{the Verifiable Fuel Cycle Simulation Model}
\newacronym{NUWASTE}{NUWASTE}{Nuclear Waste Assessment System for Technical Evaluation}
\newacronym{NWTRB}{NWTRB}{Nuclear Waste Technical Review Board}
\newacronym{OCRWM}{OCRWM}{Office of Civillian Radioactive Waste Management}
\newacronym{UFD}{UFD}{Used Fuel Disposition}
\newacronym{DYMOND}{DYMOND}{Dynamic Model of Nuclear Development }
\newacronym{DANESS}{DANESS}{Dynamic Analysis of Nuclear Energy System Strategies}
\newacronym{CAFCA}{CAFCA}{ Code for Advanced Fuel Cycles Assessment }
\newacronym{ORION}{ORION}{O..}
\newacronym{NFCSim}{NFCSim}{Nuclear Fuel Cycle Simulator}
\newacronym{COSI}{COSI}{Commelini-Sicard}
\newacronym{FCT}{FCT}{Fuel Cycle Technology}
\newacronym{SWF}{SWF}{Separations and Waste Forms}
\newacronym{FCO}{FCO}{Fuel Cycle Options}
\newacronym{RDD}{RD\&D}{Research Development and Design}
\newacronym{WIPP}{WIPP}{Waste Isolation Pilot Plant}
\newacronym{ANDRA}{ANDRA}{Agence Nationale pour la gestion des D\'echets RAdioactifs, the French National Agency for Radioactive Waste Management}
\newacronym{TSM}{TSM}{Total System Model}
\newacronym{LANL}{LANL}{Los Alamos National Laboratory}
\newacronym{INL}{INL}{Idaho National Laboratory}
\newacronym{ANL}{ANL}{Argonne National Laboratory}
\newacronym{SNL}{SNL}{Sandia National Laboratory}
\newacronym{LBNL}{LBNL}{Lawrence Berkeley National Laboratory}
\newacronym{LLNL}{LLNL}{Lawrence Livermore National Laboratory}
\newacronym{NAGRA}{NAGRA}{National Cooperative for the Disposal of Radioactive Waste}
\newacronym{CUBIT}{CUBIT}{CUBIT Geometry and Mesh Generation Toolkit}
\newacronym{CSNF}{CSNF}{Commercial Spent Nuclear Fuel}
\newacronym{DSNF}{DSNF}{DOE Spent Nuclear Fuel}
\newacronym{MTHM}{MTHM}{Metric Ton of Heavy Metal}
\newacronym{HTGR}{HTGR}{High Temperature Gas Reactor}
\newacronym{TRISO}{TRISO}{Tristructural Isotropic}
\newacronym{MA}{MA}{Minor Actinide}
\newacronym{CEA}{CEA}{Commissariat a l'Energie Atomique et aux Energies Alternatives}
\newacronym{SKB}{SKB}{Svensk Karnbranslehantering AB}
\newacronym{SINDAG}{SINDA{\textbackslash}G}{Systems Improved Numerical Differencing Analyzer $\backslash$ Gaski}
\newacronym{STC}{STC}{Specific Temperature Change}
\newacronym{LDRD}{LDRD}{Laboratory Directed Research and Development}
\newacronym{LCOE}{LCOE}{Levelized Cost of Electricity}
\newacronym{ABM}{ABM}{Agent-Based Modeling}
\newacronym{COTS}{COTS}{Commercial, Off-The-Shelf}
\newacronym{API}{API}{Application Programming Interface}
\newacronym{RIF}{RIF}{Region-Institution-Facility}
\newacronym{GUI}{GUI}{Graphical User Interface}
\newacronym{HPC}{HPC}{High-Performace Computing}
\newacronym{HTC}{HTC}{High-Throughput Computing}
\newacronym{UML}{UML}{Unified Modeling Language}
\newacronym{DAG}{DAG}{Directed Acyclic Graph}
\newacronym{XML}{XML}{Extensible Markup Language}
\newacronym{RNG}{RelaxNG}{REgular LAnguage for XML Next Generation}
\newacronym{JSON}{JSON}{JavaScript Object Notation}
\newacronym{SQL}{SQL}{Structured Query Language}
\newacronym{SQLite}{SQLite}{Structured Query Lite}
\newacronym{HDF5}{HDF5}{Hierarchical Data Format version 5}
\newacronym{CSV}{CSV}{Comma-Separated Value}
\newacronym{VL}{VL}{Variable Length}
\newacronym{SHA1}{SHA1}{Secure Hash Algorithm 1}
\newacronym{YAML}{YAML}{Yet-Another Markup Language}
\newacronym{POSIX}{POSIX}{Portable Operating System Interface}

\makeglossaries
\usepackage{color}
\usepackage{xcolor}
\usepackage{graphicx}
\usepackage{booktabs} 
\usepackage{microtype} 
\usepackage{xspace}
\usepackage{listings}
\usepackage{textcomp}
\usepackage{ulem}

\definecolor{listinggray}{gray}{0.9}
\definecolor{lbcolor}{rgb}{0.9,0.9,0.9}
\colorlet{BLACK}{black}
\colorlet{GREEN}{green}
\newcommand{\code}[1]{\lstinline[basicstyle=\ttfamily\color{green!40!black}]|#1|}

\newcommand{\cyclus}{\textsc{Cyclus}\xspace}
\newcommand{\Cyclus}{\cyclus}

\newcommand{\cycpp}{\code{cycpp}\xspace}

\date{}
\begin{document}

\begin{frontmatter}
\title{Cyclus Archetypes}

\author[usc]{Anthony M. Scopatz\corref{corrauthor}}
\cortext[corrauthor]{Corresponding Author}
\ead{scopatz@cec.sc.edu}

\author[wisc]{Matthew J. Gidden}
\author[wisc]{Robert W. Carlsen}
\author[austin]{Robert R. Flanagan}
\author[berk]{Kathryn D. Huff}
\author[wisc]{Meghan B. McGarry}
\author[wisc]{Arrielle C. Opotowsky}
\author[wisc]{Olzhas Rakhimov}
\author[wisc]{Zach Welch}
\author[wisc]{Paul P.H. Wilson}

\address[usc]{University of South Carolina, Nuclear Engineering Program, Columbia, SC 29201}
\address[wisc]{University of Wisconsin - Madison, Department of Nuclear Engineering and Engineering Physics, Madison, WI 53706}
\address[austin]{University of Texas - Austin, Department of Mechanical Engineering, Nuclear and Radiation Engineering Program, Austin, TX 78758}
\address[berk]{University of California - Berkeley, Department of Nuclear Engineering Berkeley, CA 94720}


\begin{abstract}
The current state of nuclear fuel cycle simulation exists in highly 
customized form. Satisfying a wide range of users requires model modularity 
within such a tool. \Cyclus is a fuel cycle simulator specifically designed to 
combat the lack of adaptability of previous generations of simulators. This is
accomplished through an agent-based infrastructure and treating time 
discretely. The \Cyclus kernel was developed to allow for models, called 
archetypes, of differing fidelity and function depending on need of the users. 
To take advantage of this flexibility, a user must write an archetype for 
their desired simulation if it does not yet exist within the \Cyclus ecosystem.
At this stage, a user graduates to the title of archetype developer.

Without automation, archetype development is difficult for the uninitiated. 
This paper presents the framework developed for simplifying the writing of 
archetypes: the \Cyclus preprocessor, or \cycpp. \cycpp addresses the computer 
science and software development aspects of archetype development
that can be addressed algorithmically, allowing the developer to focus on 
modeling the physics, social policies, and economics. \cycpp passes through 
the code three times to perform the following tasks: normalizing the code via 
the C preprocessor, accumulation of notations, and code generation. Not only
does this reduce the amount of code a developer must write by approximately 
an order of magnitude, but the archetypes are automatically validated.
\end{abstract}


\end{frontmatter}

\section{Introduction}
\label{sec-intro}

\Cyclus \cite{cyclus_v1_0,cyclus_v1_2} is the first truly agent-based 
\cite{jennings2000agent} fuel cycle simulator. 
New technologies, while exciting, often pose unforeseen challenges.
\Cyclus is no exception to this rule.  This paper answers the questions,
\emph{``What precisely is an agent in a fuel cycle context?''} and 
\emph{``What features of simulation can be abstracted away to ease the burden 
on fuel cycle researchers?''}

The genesis of \cyclus lies in the desire to 
treat mass balances as discrete, model facilities individually rather than as 
a fleet, and to be able to quantitatively compare the effects of changing the 
fidelity of the facilities themselves. These goals imply a large degree of 
sophistication on the part of the simulator infrastructure.  Resource exchange
must be handled in a generic and dynamic way as opposed to being hard-coded 
for specific commodities. Simulations must be inherently comparable, which involves 
storage infrastructure that is designed around this need. Agents must be able 
to communicate with one another and learn about the environment in which they 
exist. Additionally, agents must be able to be dynamically deployed and the set of 
available agent models may not be collected until run time. 

What sets fuel cycle simulation apart from traditional agent-based simulations 
is the high degree of agent specialization. Standard agent-based simulators
are characterized as having a few types of agents, often only one and 
almost always less than five \cite{taylor2014agent}. The agent types are then 
specialized 
when they are instantiated \emph{in situ}. This model is not appropriate for 
fuel cycle applications.  For example, it would be unwise to have a single facility model 
that represents both enrichment and reactors, called \code{EnrichmentOrReactor},
that decides via a switch how it behaves when it is deployed. It is much 
more natural to have two models, \code{Enrichment} and \code{Reactor}, 
that implement their own physics calculations independently.

\Cyclus enables agent specialization along two separate axes. The first defines 
an agent as an \emph{entity} that determines its role in the 
fuel cycle. There are three kinds of entities: \emph{regions} that 
represent geographic and governmental concerns, \emph{institutions} 
that manage other agents, and \emph{facilities} that implement 
physics calculations and are usually in charge of resource management.

The other axis of agent specialization distinguishes among who 
writes the model, who sets up the model for potential use in a simulation,
and who actually deploys concrete representations of the model.
At the highest level are \emph{archetypes} whose behavior is parameterizable. 
Archetypes are software implementations
of physical, chemical, economic, and political models.
For example, a \code{Reactor}  archetype may be 
parameterized by a target burnup. Authors of these highly reusable models 
are known as \emph{archetype developers}. Archetypes are in turn 
\emph{configured} into \emph{prototypes}. A prototype is a copy 
of the archetype but with all parameterizations set to concrete 
values. Hence, a \code{Reactor} with a burnup of 42 MWd/kg is a 
prototype. Configuring archetypes into prototypes is done by the \cyclus user 
in the input file. Configuration requires no underlying knowledge of 
how the archetype is implemented, though that often helps.
Finally, when prototypes are copied and \emph{deployed} in the simulation 
they become \emph{agents}. This usage of the term `agent' to mean 
the \emph{in situ} object is consistent with other agent-based
literature \cite{macal2005tutorial}.  Agent deployment happens exclusively via 
\cyclus itself; manual deployment of agents is not allowed.
Archetypes that wish to deploy agents must \emph{schedule} their building
and decommissioning.

Archetype development is one of the most difficult aspects of agent-based
simulation. Physical models be determined, implemented, and 
validated. Moreover, this is where agents interface with the 
\cyclus kernel itself. The dynamism of agent-based simulation 
coupled with the compulsion to have traditional simulator features
(such as restart and validation) creates a complex interface with the kernel.
Archetype development would be much simplified if saving and 
loading from disk and communicating with other agents were 
never a concern, though such a simulator would be of marginal use.
Due to these complexities, archetype development in \cyclus
has historically been difficult. Obtaining a working archetype
that performed no physics calculations from scratch would 
take novice developers upwards of two weeks effort.  The complexities
that lead to such a high bar are emblematic not just of \cyclus,
but of any agent-based fuel cycle simulator.

Informal polling by the \cyclus team over the years
showed that a new developer circa \cyclus v0.1
took 2+ weeks to get a working `do-nothing' archetype. This is obviously 
too long because most researchers do not have two weeks of time `just to try 
something out.' By \cyclus v0.3 the do-nothing development time had been reduced 
to approximately one business week. In \cyclus v0.4, this time became about 3 days.
As of \cyclus v1.0, this finally was reduced down to 1 - 4 hours, which meets
appropriate expectations for someone attempting \cyclus as a first time archetype 
developer.

These dramatic development time reductions were caused by two forces:
clarification of the archetype abstraction and explicit tools to help with 
archetype creation. While the region-institution-facility hierarchy was established 
early on, the formulation of this hierarchy in an agent-based paradigm took much 
longer to firmly establish.  Once this notion had been refined, attention turned to 
simplifying \Cyclus archetype development.
For a long time in the history of this simulator and its predecessors, such as 
\gls{GENIUSv2} \cite{oliver_studying_2009}, tools to help make archetypes were notable by their 
absence. 
The addition of archetype tools by the \cyclus core developers made archetype
development significantly more efficient.

The archetype development tools provided by \cyclus must overcome a variety of 
technical hurdles in order to make archetype development accessible.  These 
include, but are not limited to, the lack of reflection in the C++ programming language,
the desire to support multiple database formats, automatic validation of input files,
special mechanisms for handling resource exchange and inventory persistence, 
and the somewhat complex interface required to support the snapshotting and 
restart of simulations. Such concerns are basic to \cyclus operation but 
ancillary to the physics, chemistry, and economics being modeled by an 
archetype developer.
The minutia should `just work' since it is not a core part of a fuel cycle model.

This paper describes the strategies, efforts, interfaces,
and implementations that considerably reduce the complexity  
an archetype developer must deal with directly. This has been 
found to reduce development effort for simple agents down to a couple 
of hours for novices. Expert archetype developers realize further 
reductions down to a couple of minutes. This has has been accomplished 
without sacrificing an iota of simulation fidelity. Since such 
problems proliferate throughout all possible simulators in 
the \cyclus category, the methods described here apply beyond \cyclus.
Though such methods sometimes dive into 
computer science and agent-based modeling details, they are always implemented
with the express goal of making fuel cycle simulation 
simpler, easier, faster, more expressive, and validated.

This paper proceeds by first providing a more detailed motivation 
in \S \ref{sec-motive}. Then, \S \ref{sec-methods} describes
methods and mindsets that are used to ease archetype development.
Some of these are high-level design strategies, while others are software 
interfaces that 
provide the correct fuel cycle agent abstractions. 
\S \ref{sec-impl} describes the underlying algorithms and usages
for those methodologies, which also require concrete implementations.
Lastly, \S \ref{sec-conc} provides final remarks and illustrates
potential future directions for \cyclus archetype development.

\section{Motivations \& Problem Statement}
\label{sec-motive}

Agent-based modeling frameworks necessarily place the agent as the fundamental 
abstraction. This is true in \cyclus as well. However, many problems that are 
solved with an \gls{ABM} approach are sufficiently represented by one or two 
agents \cite{taylor2014agent}. However, in nuclear fuel cycle simulations there is
a proliferation of facility types that are distinct both conceptually and in the 
kinds of physical processes that they implement. While it would be \emph{possible}
to merge all facility types into a single, highly parameterized agent model,
it would be \emph{unwise} to do so. Facilities that model fundamentally different
physics should not be combined into a single class. Not only is it more work to 
combine them but it is also less intelligible. The same reasoning applies to why it is 
not a good idea to merge the concepts of institutions and regions with that of
facilities; the subject of the model is fundamentally distinct from other models
and this separation of concerns should be maintained.

So unlike other agent-based frameworks, the modularity of \cyclus drives 
an ecosystem of \emph{archetypes}. An archetype is an agent class that specifies 
how the agent should behave via its own implementation of physics, chemistry, 
economic, and social policies. Archetypes are parameterized only in ways that 
make sense to the policies that they implement. Extraneous policies are left to 
other archetypes. For example, a nuclear reactor would not be parameterized based 
on separation efficiencies capacity, that would be left for a reprocessing facility.

All archetypes are agents. They are able to communicate (through resource 
exchange) with all other agents and they have access to the same information 
about the environment in which they live. The archetype abstraction provides
speciation of agents so that each archetype may fill its own fuel cycle niche.

Archetypes are an essential abstraction layered on top of the agent abstraction. 
Fuel cycle facility, region, and institution modelers should directly create
archetypes rather than raw agents. Thus for people trying to create, use, and 
extend \cyclus agents, archetypes should be their entry and exit point. Because this 
concept is central to how \cyclus works, such users are known as \emph{archetype
developers}.

The archetype abstraction has the added advantage that archetype developers 
need only be specialists in the field of the archetype that they are developing.
Someone who knowledgeable about gaseous centrifuges could design 
an enrichment facility. A person that studies deep geologic repositories
could model a long-term storage facility. A reactor physicist could create 
a suite of archetypes for various reactor technologies. In this way, an ecosystem 
of archetypes from representative experts can be built up. Since one does not need
to be an expert to use an archetype, a well-developed ecosystem will provide 
a huge benefit to everyone. The separation of concerns provided by the archetype 
abstraction maximizes quality over the entire fuel cycle modeling process.

Still, for any simulator to be successful its key abstractions must be both easily 
configured by users and easily modified by developers.  If these activities are 
too difficult, the barrier to entry for new users and developers will be 
insurmountable in a reasonable time frame. Otherwise, potential new users and developers will 
walk away in confusion and frustration. For \cyclus, archetype development
needs to have first-class support.

This paper discusses how the \cyclus code base has overcome the inherent limitations
of its design goals in order 
to provide a fertile platform on which to model the nuclear fuel cycle.
Such strategies can apply to any and all agent-based fuel cycle simulators, of which 
there is currently only \cyclus.

\section{Methods \& Strategies}
\label{sec-methods}

Archetype development can be a daunting task on its own. For this reason, 
many fuel cycle simulators choose to supply only a limited corral of 
pre-built archetypes. This places the responsibility of creating an maintaining 
the archetypes with the authors of the simulator.  When only pre-built archetypes are available, then this suite defines the scope 
of all possible fuel cycles that a user can model. If the scope is not broad 
enough for the needs of a user, then the simulator developer must expand the 
scope or lose the user.  This can create a bottleneck because the number of users 
may increase while the productivity of the simulation developer remains constant.

\Cyclus avoids such bottlenecks by empowering users to create and maintain 
their own archetypes independent of the development timeline of the \cyclus 
kernel. However, this modularity comes with its own costs. Being able to 
plug in user-created archetypes implies an \gls{API} to which 
the archetype conforms. This is an 
additional burden to both nascent and experienced archetype developers that 
does not effect the underlying behavior or physics. Rather, a large percentage of the
code for an archetype exists soley to satisfy the \cyclus \gls{API} and does not
impact the physics, economics, or other domain concerns being modeled.

To mitigate the difficulties of writing archetypes, 
\cyclus must aid in overcoming the hurdle of interfacing with 
kernel itself. Additionally, to the extent possible, \cyclus should 
also provide tools that 
ease the implementation of the physics and desired behavior. 
A variety of strategies are used by \cyclus to ease archetype development: 

\begin{itemize}
    \item \textbf{Automatic Model Templating:}  By automatically 
        inspecting and creating portions of archetypes, the overhead
        of adhering to the \cyclus \gls{API} is removed. This is performed 
        via preprocessing archetypes and then applying limited code generation
        to the original model. Such activities also 
        add limited reflection to the C++ models as needed, which is important 
        for archetypes that wish to know about themselves.

    \item \textbf{Data Communication Protocol:} \Cyclus 
        provides a common basis for archetypes to store and retrieve 
        complex data in the database. This is achevied through the 
        implementation of \Cyclus-specific type system.
        This alleviates the need for archetype
        developers to invent and implement custom persistence solutions.

    \item \textbf{Validation:} Archetypes use \gls{XML} schema to validate that 
        their prototypes have been configured correctly.
        This ensures the users of an archetype are adhering to the contraints
        imposed by the developer.

    \item \textbf{Metadata Annotations:} Archetypes have a standard place to 
        store and retrieve both pre-defined and arbitrary metadata.
        This allows for archetype developers to communicate relevant information
        to tools outside of \cyclus (such as a visualization tool).

    \item \textbf{Model Location:} \Cyclus has a packaging system for archetypes and 
        libraries of archetypes. This provides a standard mechanism for 
        searching for and locating models. 
        Furthermore, all archetype developers can 
        uniquely specify their own archetypes without the fear of overlapping 
        names.  For example, two developers could each have a \texttt{Reactor}
        archetype, but they would exist in different packages and be 
        disambiguated.

    \item \textbf{Markets are not Agents:} In an agent-based methodology, 
        the mechanism for communication between agents is not an agent itself.
        Thus in \cyclus, market resolution was moved to be solely in the purview 
        of kernel. In order to transfer resources, agents communicate 
        in a well-defined way. Thus, archetype development need not  
        include markets that specify how archetypes wish to communicate.

\end{itemize}

Though the above mechanisms are discussed with respect to \cyclus, they are 
transferable to any agent-based modeling framework that requires modular agent 
archetypes. The following subsections present greater detail regarding these 
strategies.

\subsection{Automatic Model Templating}
\label{subsec-ppgc}

Every \cyclus archetype is required to implement the member functions 
seen in Table \ref{req-api} and may optionally implement those seen in 
Table \ref{opt-api}. Due to object orientation in 
C++ and how \cyclus stores state, these member functions must be implemented directly
on the archetype itself. The implementation of these functions that archetypes 
inherit from the \code{Agent} class is not and cannot be sufficient.

\begin{table}
\caption{Required Archetype Interface}
\label{req-api}
\begin{tabular}[p]{|lp{5.25in}|}
\hline
\textbf{Function} & \textbf{Description} \\
\hline
\code{InfileToDb()} & Reads the prototype in the input file (XML format) 
                      and adds them to the initial startup database.\\
\code{InitFrom(Db)} & Initializes a new agent or prototype from the database.\\
\code{InitFrom(Agent)} & Iniializes a new agent or prototype from another agent or
                         prototype.\\
\code{InitInv()} & Initializes any starting inventory buffers.\\
\code{Clone()} & Copies the current agent.\\
\code{Snapshot()} & Stores the current state of the agent in the database.\\
\code{SnapshotInv()} & Stores the current inventory buffers in the database.\\
\code{schema()} & Returns the schema that user input must validate against.\\
\code{annotations()} & Returns all metadata that is automatically gathered
                       and supplied by the archetype developer.\\
\hline
\end{tabular}
\end{table}

\begin{table}
\caption{Optional Archetype Interface}
\label{opt-api}
\begin{tabular}[p]{|lp{5.25in}|}
\hline
\textbf{Function} & \textbf{Description} \\
\hline
\code{Build()} & Called when the agent has been built.\\
\code{EnterNotify()} & Allows for the agent to register for services.\\
\code{BuildNotify()} & Informs when children of this agent have been built.\\
\code{DecomNotify()} & Informs when children of this agent are about to be 
                       decommissioned.\\
\code{Decommission()} & Removes the agent from the simulatuon.\\
\hline
\end{tabular}
\end{table}

Archetypes store state as public or private member variables
directly on the class.  For example, a \code{Reactor} archetype class
could have a burnup declared as `\code{double burnup};'. Not every agent has 
a burnup (e.g., an enrichment facility) and so the \code{burnup} member 
should not be part of \code{Agent}. Furthermore, archetypes are modularly 
defined. Because \cyclus cannot know all possible field names 
for all possible archetypes for all time, it can only react to how 
archetypes are written. In a dynamic language, the \code{Agent}
superclass would still be able to to \emph{inspect} instances of its subclasses
to discover field names at runtime. Such introspection is called \emph{reflection}. 
This would allow \code{Agent} to hold a single, generic implementation of each of the
member functions in Listing \ref{req-api}, relieving archetype developers from 
having to implement them manually.

However, C++ lacks reflection. Neither \code{Agent} nor archetypes are allowed
to dynamically discover what their member variable names are at runtime.  This 
implies that the archetype developer must explicitly code in the appropriate variable
names in the correct way for each of the nine required functions.  For instance, 
the \code{Reactor} class must explicitly save the \code{burnup} member in the 
\code{Snapshot()} function if \code{burnup} is to be saved to the database.
None of the nine required interfaces comes along automataically and minor typos 
could cause large breakages. For example, misspelling \code{burnup} as \code{bunrup}
could cause the \code{Snapshot()} method to fail silently if it occured in the 
wrong location.

Since implementing this part of the archetype interface is both highly 
error-prone and 
routine, it is ripe for automatic code generation. Code generation replaces the 
tedious task of writing the required member functions with software that will 
insert such member functions into an otherwise fully developed archetype. 

The code generation strategy has some limitations. 
First, it must be 
performed to prior to compilation. Second, the
code generator must be provided with enough information about the archetype in 
order to accurately create the function implementation. Third and finally, it 
is highly desirable to keep archetypes in valid C++, rather that a special 
code generation language created just for \cyclus.
Templating languages 
such as Jinja \cite{ronacher2011jinja2} or any other variety of custom 
solutions would allow for 
expressive code generation. However, such template languages would be yet another 
tool for the archetype developer to learn. This runs counter to the goal of 
simplifying development.

The limitations above are all elegantly addressed through the use of a 
\cyclus-aware preprocessor. The first stage of C/C++ compilation is the 
C Prepocessor or \code{cpp} \cite{stallman1987c}. This tool is responsible for expanding 
\code{#include} directives and implementing other \code{#} directives, such as 
\code{#pragma}. It is executed
prior to any other stage of compilation (lexing, parsing, \emph{etc.}).
Importantly, the \code{#pragma} directive is skipped by \code{cpp} if it is not 
recognized and the directive is passed though to further preprocessors or a C++ 
compiler. It is a
purposeful hook for other preprocessors to use and implement their own code generation.
If an alternative preprocessor only uses \code{#pragma} directives as its interface, 
the developer will be able to write in pure C++ and reap the benefits
of an extra code-generation step. 

However, all code generators must be supplied with sufficient information about
where and how to create the code. These tasks may also be accomplished through 
the use of pragmas. To handle a suite of such utilities, a custom preprocessor
is needed.

Certain pragmas may be used to parse only the needed information 
about an archetype, rather than parsing the entire class. For an archetype, the 
member variables that are saved and loaded from its instances are the most important.
This is because they fully describe the state of an agent 
at all points in the simulation. These member variables are known as \emph{state variables}. 
To generate the appropriate input and output routines for an archetype, at a minimum, the names and C++ types 
of all state variables must be known. The gathering of all these state variables is is 
called \emph{state accumulation}. 

Still, other pragmas denote where to insert automatically generated code into the 
original file. This is the tedious 
part of achetypes communicating with the \cyclus kernel. Specifically, 
the member functions in Listing \ref{req-api} ought to be created and inserted 
for the
archetype developer, avoiding entirely the need to write them by hand.

The \cyclus preprocessor, called \cycpp, handles all of the cyclus code generation
and state accumulation. This tool only recognizes directives that begin with \code{#pragma cyclus} so as to uniquely distinguish it from other preprocessors.
\cycpp performs three complete passes through archetype code:
\begin{enumerate}
    \item Normalization of the source via the standard C preprocessor \code{cpp} 
    \item Accumulation of state and other agent annotations from normalized code 
    \item Code generation into original source
\end{enumerate}
Importantly, any preprocessor that adds reflection must make at
minimum two passes: discovery of what exists (state accumulation) and adding this
information back to the class (code generation). Single-pass preprocessors such 
as the standard \code{cpp} utility are not sufficient to enable reflection because they 
cannot guarantee that all relevant class information has been seen when code generation begins. 
Thus, \cycpp includes these two passes plus an additional 
initial pass to simplify state accumulation. Further passes could be added that 
implement reflection onto the generated code itself, but this is often unnecessary. 
Particularly in \cyclus, further passes would be excessive since the portions of the 
\code{cyclus::Agent} interface that are generated are known ahead of time.

\cyclus-specific pragmas are broken up into two categories depending on whether 
they are most relevant to the state accumulation or code generation steps. They 
are denoted as \emph{annotation directives} and \emph{code generation directives},
respectively.

The \cycpp annotation interface has two main directives:
\begin{itemize}
    \item \code{#pragma cyclus var <dict>} - state variable annotation
    \item \code{#pragma cyclus note <dict>} - agent annotation or note
\end{itemize}
The state variable annotation is used on the line immediately above an archetype
member variable to declare it as a state variable. The note directive is used
anywhere in an archetype class declaration and applies to the archetype itself.
Both of these have a \code{<dict>} argument that is a Python dictionary. This 
holds metadata about the state variable or the archetype. For example, a \code{flux}
state variable declaration is shown in Listing \ref{flux-pragma}.

\begin{lstlisting}[caption={Flux State Variable Annotation}, label=flux-pragma]
#pragma cyclus var {"default": 42.0, "units": "n/cm2/s"}
double flux;
\end{lstlisting}

The code generation interface in its simplest form occurs via the \cyclus 
prime directive, \code{#pragma cyclus}.  This expression will 
generate the entire archetype interface and insert it in place of the pragma.
More fine-grained code generation may be used by passing additional arguments.
The signature for such targeted code generation may be seen in Listing \ref{targ-cg}.
The first argument is one of \code{decl}, \code{def}, or \code{impl} representing 
interface declarations, definitions (the declaration and the implementation together), 
or implementations (just the body without the declaration) of the archetype member
functions respectively. Lacking any further arguments, this will produce the desired code for 
all of member functions in Listing \ref{req-api}.  Optionally, a member function 
name in all lowercase (i.e., \code{inittodb} rather than \code{InitToDb}) may 
be supplied to generate code for the given function alone.  Lastly, in 
cases where \cycpp cannot determine the archetype to apply the code generation for,
such as in an implementation file (\code{*.cc} or \code{*.cpp}), the agent
class name may be provided as a final parameter.

\begin{lstlisting}[caption={Targeted Code Generation Directive Signatures}, 
                   label=targ-cg]
#pragma cyclus <decl|def|impl> [<func> [<agent>]]
\end{lstlisting}

When the annotation and code generation directives are combined and expanded with 
\cycpp, agents become much simpler to write.
Archtype developers are then free to focus on their physics 
or economics algorithms. The base \cyclus
infrastructure is largely removed from the concern of the archetype developer.
For example, a simple reactor model may be completely implemented as seen in 
Listing \ref{rx-eg}. The barrier to creating new archetypes is much lower than 
compared to the hundreds of lines of code that this would require without \cycpp.

\begin{lstlisting}[caption={Simple Reactor Archetype}, label=rx-eg]
class Reactor : public cyclus::Facility {
 public:
  Reactor(cyclus::Context* ctx) {};
  virtual ~Reactor() {};

  #pragma cyclus

 private:
  #pragma cyclus var {'default': 4e14, 'units': 'n/cm2/2'}
  double flux;

  #pragma cyclus var {'default': 1000, 'units': 'MWe'}
  float power;

  #pragma cyclus var {'doc': 'Are we operating?'}
  bool shutdown;
};
\end{lstlisting}

\subsection{Data Communication Protocol}

Related to the issue of how to snapshot and restart simulations is the issue of 
what fundamental data types are allowed to be saved and loaded natively by the 
simulation. For many physics simulators, primitive data types (i.e., \code{float}, 
\code{int}, and \code{string}) and arrays of these types are sufficient to 
express and evaluate the underlying equations. This handful of types is small enough 
that most simulators can handle them in an ad hoc manner. Furthermore, these types
often align with the structure of most databases, allowing for easy translation
between disk and memory.

Agent-based fuel cycle modeling, however, subsumes physical modeling frameworks.
The types of data that are needed to naturally express many fuel cycle concepts 
do not fit into the standard `arrays of floats' mindset.  More sophisticated types are needed,
such as maps and sets of primitive types. For example, 
a reprocessing facility needs to have a mapping from elements to separation 
efficiencies. To avoid hard-coding the elements into the archetype, the user should 
be able to state which elements are allowed for each separation. The correct data
type in C++ for such information is \code{std::map<int, double>}. Any other 
representation, say an \code{int array} or a \code{float array},  would likely be converted to 
a \code{map} by the archetype itself.

\Cyclus increases expressiveness of archetypes and reduces error from extraneous 
transcription by natively supporting its own extensible type system. Every type
has its own unique integer identifier as well as 
a variable name corresponding to this type. 
The types of the state variables are given by corresponding C++ representations.
Every database format may then choose which types in the type system it 
supports and how it implements them. All types have a static \emph{rank}, or the 
number of variable length dimensions, they support.  For example, \code{int} and 
\code{double} are both rank-0, while \code{vector<float>} is rank-1 due to 
vectors having arbitrary length. As an optimization, the archetype developer 
may also give a \emph{shape}, or the maximum size along each dimension. 
The type system is extensible, 
allowing for expressiveness to grow and evolve with the needs of archetype developers.

This strategy represents a significant abstraction over the needs and usage of most 
other simulator technologies. This level of detail is required by \cyclus due to the
dynamically loadable agents. Without a strong type system, state variables 
would only be minimally useful and archetypes would have to rely on out-of-simulator
mechanisms to save and load their state.

\subsection{Metadata Annotations}

Metadata annotations are coupled with code generation.  State variable annotations
are provided by using the \code{#pragma cyclus var} directive. Annotations for 
the archetypes as a whole are given using the \code{#pragma cyclus note} directive,
as specified in \S \ref{subsec-ppgc}. Both of these directives take a dictionary 
as an argument and this mapping must have string keys.

Metadata serves an important purpose by communicating information to the \cyclus
preprocessor, to the \cyclus kernel, and even beyond the simulation to analysis 
and visualization tools. Some metadata is automatically generated from the 
archetype declaration itself.  This information is considered read-only and required. 
Other annotations are supplied by the archetype developers and are optional. 
Some keys, such as documentation, are highly recommended even though they 
are optional.

Any key may be supplied to the annotation directives. For most keys, the 
purpose of the entry is given entirely by the archetype developer. However, 
some keys are reserved and have special meanings in various contexts. Table
\ref{sv-anno} lists these keys for state variable annotations and Table
\ref{ag-anno} displays the reserved keys for \code{#pragma cyclus note}.

\begin{table}
\caption{Special State Variable Annotations}
\begin{tabular}[htbc]{|p{0.2\linewidth}|p{0.75\linewidth}|}
\hline
\textbf{key} & \textbf{meaning} \\
\hline
\code{type}    & The C++ type. \emph{Read-only}.\\
\hline
\code{index}   & State variable order-of-appearence, 0-indexed. \emph{Read-only}.\\
\hline
\texttt{default} & The default value for this variable that is used if otherwise 
                 unspecified by the user. The type of the value must match the 
                 type of the variable.\\
\hline
\code{shape}   & The shape of the data type. If present this must
                 be a list of integers of a given length (rank).
                 Specifying positive values will, depending on the 
                 backend, render this a fixed-length data type
                 of the provided length. A value of \code{-1}  
                 will retain the variable-length along that axis. 
                 Fixed-length variables are normally more performant and thus it is 
                 often better practice to specify a shape if possible. For 
                 example, a length-5 string would have a shape of \code{[5]} and 
                 a length-10 vector of variable-length strings would have a 
                 shape of \code{[10, -1]}.\\
\hline
\code{doc}     & Documentation string.\\
\hline
\code{tooltip} & Brief documentation string for user interfaces.\\
\hline
\code{units}   & The physical units, if any, as a string.\\
\hline
\code{userlevel} & Integer (0 - 10) representing ease (0) or difficulty (10) 
                   in using this variable, default 0.\\
\hline
\code{schematype} & The data type that is used in the schema for input file
                    validation. This enables the user to supply just the data type
                    rather than having to overwrite the full schema for this state
                    variable. In most cases - when the shape is rank 0 or 1 such
                    as for scalars or vectors - this is simply a string. In cases
                    where the rank is 2+ this is a list of strings. Please refer to
                    the \gls{XML} Schema Datatypes \cite{xml-datatypes}
                    for more information.\\
\hline
\code{initfromcopy} & Code string to use in the \code{InitFrom(Agent* m)} 
                      function for the state variable instead of automatic code 
                      generation.\\
\hline
\code{initfromdb} & Code string to use in the 
                    \code{InitFrom(QueryableBackend* b)} 
                    function for this state variable instead of automatic code 
                    generation.\\
\hline
\code{infiletodb} & Code strings to use in the \code{InfileToDb()} function 
                    for this state variable instead of automatic code generation.
                    This is a dictionary of string values with the keys `read'
                    and `write' that represent reading values from the input file 
                    writing them out to the database, respectively.\\
\hline
\code{schema}  & Code string to use in the \code{schema()} function for 
                 this state variable instead of automatic code generation.
                 This is an \gls{RNG} string. This is usually coupled with 
                 \code{infiletodb} to ensure the custom
                 schema is read into the database correctly.\\
\hline
\code{snapshot} & Code string to use in the \code{Snapshot()} function for 
                  this state variable instead of automatic code generation.\\
\hline
\code{snapshotinv} & Code string to use in the \code{SnapshotInv()} function 
                     for this state variable instead of automatic code generation.\\
\hline
\code{initinv} & Code string to use in the \code{InitInv()} function for 
                 this state variable instead of automatic code generation.\\
\hline
\end{tabular}
\label{sv-anno}
\end{table}

\begin{table}
\caption{Special Agent Archetype Annotations}
\begin{tabular}[htbc]{|p{0.2\linewidth}|p{0.75\linewidth}|}
\hline
\textbf{key} & \textbf{meaning}\\
\hline
\code{vars} & The state variable annotations. \emph{Read-only}.\\
\hline
\code{name} & C++ class name (string) of the archetype. \emph{Read-only}.\\
\hline
\code{entity} &  String of the type of archetype based on which class it 
                 inherits from; \code{cyclus::Region},
                 \code{cyclus::Institution}, or \code{cyclus::Facility}are
                 given by `region', `institution', or `facility',
                 respectively. If the class inherits from \code{cyclus::Agent} but 
                 not the previous three 
                 the string `archetype' is used. The string 'unknown' is used
                 if the class does not inherit from \code{cyclus::Agent}.
                 \emph{Read-only}.\\
\hline
\code{parents} & List of string class names of the direct super-classes of this
                 archetype. \emph{Read-only}.\\
\hline
\code{all_parents} & List of string class names of all the super-classes of this
                     archetype. \emph{Read-only}.\\
\hline
\code{doc} & Documentation string.\\
\hline
\code{tooltip} & Brief documentation string for user interfaces.\\
\hline
\code{userlevel} & Integer (0 - 10) representing ease (0) or 
                   difficulty (10) in using this variable, default 0.\\

\hline
\end{tabular}
\label{ag-anno}
\end{table}

With respect to \cycpp, metadata enables the customization of code generation
without the need to alter the preprocessor itself or create an entirely new 
code generator. This tailoring of \cycpp is exemplified by keys, such as \code{'shape'}, 
\code{'schematype'}, and \code{'initfromcopy'}.  Still, other keys are generated by \cycpp itself
and are considered read-only.
These include the all-important \code{'type'}, \code{'index'}, \code{'name'}, and 
\code{'parents'} keys.  The automatic
creation of these keys minimizes human transcription error for the most important 
metadata. This is a core simplification of archetype development.

Metadata is similarly important for the \cyclus kernel. All metadata is 
directly available via the \code{annotations()} member function, which returns 
a \gls{JSON} 
object (equivalent to a Python dictionary with string keys). Combined with the 
auto-generated read-only keys from \cycpp, the metadata provides much needed 
reflection to the archetypes.  Unlike most C++ classes, archetypes have 
runtime access to their own class names, their parent classes, and the names and 
types of their state member variables. Therefore, the archetype may 
make runtime decisions about how to behave based on how it is defined.
The major current use for the limited reflection in archetypes is for agents to 
save and load themselves. 

Lastly, metadata is useful beyond \cyclus itself. Annotation keys such as \code{'doc'},
\code{'tooltip'}, and \code{'units'} are included to provide end-user documentation. 
The \code{'userlevel'}
and other keys signal how archetypes and state variables should be treated in 
downstream user interfaces 
Even keys, such as \code{'default'}, 
that might have a primary usage elsewhere
may still be helpful beyond the scope of \cyclus kernel.  

The \code{#pragma cyclus var} and \code{#pragma cyclus note} directives provide 
unambiguous locations for implementing and 
generating metadata annotations. Metadata being contained completely within archetype 
declarations increases the worth of both the metadata and the archetypes
because there is a single source for information about archetypes.
Additionally, this high degree of locality eases the creation of
archetypes, enabled by the automatic discovery of the annotation keys.

\subsection{Validation}

When writing \cyclus input files, users configure archetypes into prototypes.
Configuration is done by assigning all state variables to 
initial values. This is the principal mechanism by which information flows 
from users to archetype developers. Therefore, it is reasonable for archetype 
developers to ask, ``Does what the user gave me make sense?'' For example: 

\begin{itemize} 
    \item a flux state variable should not be negative
    \item an integer variable should not receive the string ``Toaster''
    \item a vector representing an N-group cross-section should have exactly N 
          elements
\end{itemize} 
If the user were to not follow these rules, the input file would be break physical
constraints, type constraints, or shape constraints, respectively. 
Such breaking of constraints must be an error and the simulation should not be 
allowed to run. 
These rules can be codified into the archetypes allowing 
the simulator to check that whether or not any input file adheres to the 
given constraints. This is known as \emph{validation} and \cyclus implements 
it automatically to the extent possible. Input file validation occurs prior
to any simulation.

\Cyclus input files are written in \gls{XML}, which  may be validated against a 
known structure called a \emph{schema}. The schema is itself \gls{XML} that describes 
the layout, types, and attributes of the input it will validate.
\Cyclus provides a default schema that describes the overall structure of the 
input file.

Archetypes, however, are dynamically loaded and so schema for their state 
variables may not be predicted or preloaded. To ensure consistency, archetype schema 
must be loaded along with the archetype itself.  It may be tempting to ignore the 
notion of archetype schema and not validate the state variables. 
Howevere, this would
create a system where there is no contract between the 
user and the archetype developer. Even if an archetype developer implemented 
\emph{ad hoc} validation to their own classes, users would not
anticipate such 
restrictions.  To avoid this, \cyclus requires
that archetypes provide their own validation via the \code{schema()} member 
function seen in Listing \ref{req-api}.

To accomplish the
above, the schemas must be written in a schema language, which happens to be a 
subset of \gls{XML}. Thus the schema interface is two steps removed from the C++ that 
defines the archetypes. 
This provides additional 
cognitive load to the archetype developer as they now must learn two additional
tools before writing an archetype. However, the schema for most state
variables may be derived automatically from information known to the preprocessor:
the name of the state variable, its type, and optionally its shape and size in 
the case of containers. Thus the \code{schema()} member function is auto-generated
by \cycpp and input validation is obtained for free. 

\gls{XML}-based schema are extraordinarily useful as a mechanism for validating types. 
For example, if the user provides a floating point number rather than an 
expected string, they should be alerted to this error immediately.
Moreover, schema can also validate structure to assert that 
a variable has the right shape. A length-5 vector must be initialized with 
five elements. 

However, semantic and physical meaning must be ascribed by the archetype developer.
Giving meaning to state variables based
on their name and C++ type alone is impossible to accomplish via an automated
method. For example, that a variable named \code{flux} 
on a \code{Reactor} should not be negative comes from a physical understanding
and not a computational one. Such meaning can and should be given by the 
archetype developer via metadata.

Metadata for physical validation amounts to modifications of the schema. There are 
two annotation keys that may be used to accomplish this.  The first clarifies the 
type that the schema uses and is called \code{'schematype'}.  
In the example of a group structure shown in Listing \ref{ngroups}, the 
\code{'schematype'} could be assigned a value of 
\code{'positiveInteger'} rather than relying on the default \code{'int'} type
that permits negatives and zeros. Additionally, the metadata key \code{'schema'}
can be used to wholly replace the auto-generated schema for the state variable at 
hand. This key can be used to change the name of the state variable with respect 
to the schema or to make a variable optional.

\begin{lstlisting}[caption={Physical Constraint Addition via `schematype'}, label=ngroups]
#pragma cyclus var {'schematype': 'positiveInteger'}
int ngroups;
\end{lstlisting}

Finally, \Cyclus also allows for the circumvention of automatic schema generation. This 
provides another method for instilling semantic meaning into state variable annotations
that does not rely on metadata annotations, although
partial or complete circumvention of code generation will require
more effort from the archetype developer. The fine-grained control afforded by 
hand-writing the \code{schema()} member function is performed 
by advanced developers and only when absolutely necessary due to insufficiencies 
in \cycpp.

\subsection{Model Location}

In a robust ecosystem of archetypes, it is nearly guaranteed that different archetype
developers will want to use the same name. No single person or organization can 
reasonably lay sole claim to generic terms such as \emph{reactor}, \emph{source}, 
\emph{sink}, and other names. Simultaneously, the archetype developer should not 
be concerned with accidental name collision between their archetypes and archetypes
of other
past, present, and future developers.  Furthermore, it is often useful in a 
simulation or 
development campaign to group similar or related archetypes together. Uniqueness
and collection problems are simultaneously solved through a hardy \emph{package system}.

\cyclus packaging is an organizational structure that defines where on the file system 
archetypes are installed to, how the \cyclus kernel will load installed
packages, and how to uniquely identify an archetype in an input file.  Archetypes 
are denoted with a three-part \emph{archetype specification}. When spelled out, this
is a colon-separated string with the following elements:
\begin{enumerate}
    \item a slash-separated (\texttt{/}) directory path,
    \item a library name, and
    \item an archetype name.
\end{enumerate}
For example, \code{my/path:mylib:MyAgent} represents \code{MyAgent} living
in \code{mylib} residing in the \code{'my/path'} directory. More than just 
a simple spelling convention, this is a necessary
tool for searching for and discovering archetypes on the machine of the user.

The path portion of the specification is relative to the \code{CYCLUS_PATH}. This is 
an list of directories on the machine that will be searched in order to find the 
archetype of interest. By default, \code{CYCLUS_PATH} contains the current working 
directory, the \cyclus install directory, and the \cyclus build directory. 
\code{CYCLUS_PATH} may also be modified as an environment variable, allowing the user
to permanently or temporarily alter the \cyclus search behavior.  Thus the path 
specification (e.g. \code{'my/path'}) is interpreted as a sub-directory of all of 
the directories on the \code{CYCLUS_PATH}. For a directory \code{d1} on 
\code{CYCLUS_PATH}, if  \code{'d1/my/path'} does not exist then the search for 
the archetype will continue along with \code{d2}, and so on. The path portion 
may be an empty string, indicating that the library lives directly on the 
\code{CYCLUS_PATH}.

The library name is the dynamically loadable library file name that stores the 
archetype. This does not include the \code{lib-} prefix or the file extension, 
which is generally operating system-dependent.  For example, on a
POSIX system,
a file named \code{libmyagents.so} would receive the library name \code{myagents}
in the archetype specification. 
If a library name is not specified, then it is assumed to be the same as the 
archetype name.
If desired, a single path may hold many libraries and a single library may hold
many archetypes. Thus, archetypes may be grouped
together coarsely or finely, depending on the needs of the archetype developers.

The path and library names together allow for complete disambiguation of 
archetypes because they enforce an important degree of namespacing.  It is unlikely
that two well-designed libraries will overlap in both library and archetype name.
Even if they overlap, one or both libraries may be placed in respective
sub-directories and the path is used to establish uniqueness.
This strategy for specifying archetypes ultimately removes confusion and error from 
both archetype developers and users alike.

\subsection{Markets are not Agents}

In early versions of \cyclus, the dynamic resource exchange algorithm that the 
kernel used was itself dynamically loadable. Such an algorithm was called a
\emph{market} and was categorized as an entity alongside regions,
institutions, and facilities. 
Each commodity was traded in its own market, which was specified by 
the user in the input file.

Unlike the other entities, though, a market did not have agency.  It could not 
communicate with other agents in the simulation because it was itself the method
for agent communication.  Furthermore, resource exchange is the keystone
part of all fuel cycle simulators. 
Relegating such algorithms to live outside of 
the kernel lead to maintenance problems. It became difficult to 
ensure that all markets correctly supported the proper exchange interface,
thus minimizing the value of modularity within \cyclus.

Therefore, the notion of markets as a simulation entity was removed. In their 
stead, dynamic resource exchange algorithms were brought into the core to
guarantee exchange feasibility. Moreover, this enabled all commodities to trade 
through a single global exchange. The commodity itself automatically defines 
the sub-exchange graph in which the commodity participates.
These sub-exchange graphs may be thought of as analogous to the 
markets, which were removed in order to simplify archetype development.
Since markets did not initially have any agency, their removal did not 
affect the agent-based
nature of \cyclus.  Rather, market removal allowed archetypes to 
communicate through a common resource exchange interface.

In summary, the current interpretation of dynamic resource exchange 
eases the burden on archetype developers. Because the primary duty of the 
kernel is to provide generic and valid resource exchange algorithms,
the archetype developer is not required to construct a custom exchange for 
each commodity an archetype trades. Furthermore, market removal from the
kernel does not 
impinge on exchange solver availability or customization.  Many exchanges may 
be provided via user-tuneable parameters.  The only restriction 
is that the exchange algorithms must exist within \cyclus itself.  This is not 
considered overly burdensome, because individuals seeking to write custom 
exchanges - arguably the most advanced task in \cyclus - have likely 
transitioned from being an archetype developer to also being a kernel developer.

\section{Implementation}
\label{sec-impl}

In \S \ref{sec-methods}, the strategies and interfaces that \Cyclus uses to 
simplify archetype development were presented. These represent notions about
the amount of information and prior knowledge that the archetype developer 
must have in order to write archetypes.  If a particular strategy 
decreases the knowledge required by archetype developers then it is considered
beneficial to implement.  

However, methods that are more intuitive for new users to understand are often
proportionately more difficult to implement. For example, playing or mastering 
the game \emph{tic-tac-toe} is a vastly different effort than
designing the game in the first place.  
This section describes the infrastructure of
current \cyclus archetype development.  This is relevant to other fuel 
cycle simulators that wish to adopt the same strategies that \cyclus 
implements. In particular, the implementation of the \cyclus preprocessor, 
the type system, input file validation, and metadata annotations will all 
be covered here.

\subsection{The \Cyclus Preprocessor}

The \cyclus preprocessor, \cycpp, is responsible for all metadata collection and 
code generation for archetypes. It is implemented as a small Python utility 
that is currently less than 2000 lines in a single file.  It has no dependencies other 
than the Python standard library \cite{lutz2010programming}. It is thus light-weight enough to move around 
between code projects, if needed. For the scale of its responsibility, \cycpp
is extremely efficient. 

The preprocessor implements the three passes detailed in \S\ref{subsec-ppgc}:
normalization via standard \code{cpp}, state variable annotation accumulation, and code 
generation. The \cycpp tool must be run on all C++ header and source files that
contain archetype code and the \code{#pragma cyclus} directives. Running \cycpp
on files without such directives will result in no changes to the 
original file. The first \cycpp
pass that runs the C preprocessor is a trivial subprocess 
spawn. Importantly, this ensures that \cycpp detects the
same include, macro definitions, and macro un-definitions that actual 
compiler will see.

The second pass, state accumulation, represents half of the work that \cycpp performs.
The results from pass one are fed into this pass and scoured for 
potentially relevant information about the archetypes present in the file. 
Thus, state accumulation 
may be thought of as a traditional parser that transforms tokens (lines of the 
C++ file) into a more meaningful in-memory data structure. As a parser, pass two 
may be implemented as a \emph{state machine}
\cite{mertz2003text,wagner2006modeling}.

\begin{figure}[htbc]
\centering
\includegraphics[width=0.8\textwidth]{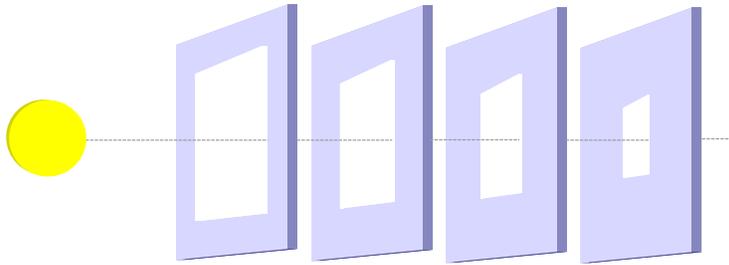}
\caption{The \code{StateAccumulator} class passes lines of C++ code through 
a series of filters, each of which may transform the information heretofore gathered
by previous filters. 
This may be thought of analogous to a spheres of various radii traversing 
concentric windows.  The spheres, or lines of code, stop when they cannot pass a  
window, or filter. This triggers the execution of the transformation function of just 
that filter.}
\label{filter-analogy}
\end{figure}

Pass two is represented in \cycpp by the \code{StateAccumulator} class,  a
state machine that compares the output lines from the C preprocessor against a series of \emph{filters}.  If a line matches the expected structure 
for a filter, then the filter executes a \emph{transformation} function on the 
line and no further filters are executed. If the line does not match any 
filters, then the line is allowed to pass through the \code{StateAccumulator}.  
The filter-transformation sequence can be thought of in analogy to a sphere of 
a given radius (a line of code) attempting to pass through concentric windows 
(the filters) of decreasing aperture, as illustrated in Figure \ref{filter-analogy}. 
This first window where the sphere stops 
represents the transformation that is executed.  This sphere is allowed to move 
through the system without being stopped.

The filters implemented for pass two of \cycpp are described in Table \ref{pass2-filters}, 
in order of decreasing precedence. The most important of these filters implement 
the \code{#pragma cyclus} directives that the archetype uses to communicate
with the preprocessor.  The pragma filters typically modify attributes of
the \code{StateAccumulator}, such as the \code{context}, the \code{execns} 
(or \emph{execution namespace}), the \code{aliases} set, and the \code{namespaces}. 
These represent the classes, types, aliases, and other information that defines
the scope of the C++ code. Such information is necessary 
for accurately representing archetypes and their state variables. 

Most pass two filters that do not implement a preprocessor directive instead
aggregate information about the available types. The \code{VarDeclarationFilter} has
the important job of determining the C++ type of state variables from the 
member variable declaration on the archetype class. However, the C++ type system 
is complex and allows for a number of programmer modifications prior 
to the type declarations:
\begin{itemize}
    \item types may be aliased to any number of alternative names, 
    \item template types are white-space insensitive, 
    \item scoping rules apply to new type names, and 
    \item other issues must be resolved to accurately and uniquely represent a C++ type. 
\end{itemize}
This requires that \cycpp implement relevant type handling
simply to correctly spell the canonical type name. 
Thus the \code{StateAccumulator} class acts as its own type system and
returns the canonical form of any type it knows about at all points during 
pass two of \cycpp.

\begin{table}
\caption{\cyclus Preprocessor Pass 2 Filters (Higher order filters have 
         lower execution precedence)}
\begin{tabular}[htb]{|p{0.05\linewidth}|p{0.33\linewidth}|p{0.6\linewidth}|}
\hline
\textbf{order} & \textbf{filter} & \textbf{description} \\
\hline
1  & \code{ClassAndSuperclassFilter} & Accumulates the class name from a class 
                                       declaration. Also stores the names of the 
                                       superclasses from the declaration.\\ 
\hline
2  & \code{AccessFilter} & Sets the current access control level:
                           \code{public}, \code{private}, or \code{protected}.\\
\hline
3  & \code{ExecFilter} & Implements the \code{#pragma cyclus exec <code>} directive
                         that allows for the execution of arbitrary Python code.
                         The results of this code are added to the context that 
                         evaluates other \cycpp directives.\\ 
\hline
4  & \code{UsingNamespaceFilter} & Adds and removes a namespace from the 
                                   current scope via the C++ \code{using namespace}
                                   statement.\\ 
\hline
5  & \code{NamespaceAliasFilter} & Implements namespace aliasing in the current 
                                   scope.\\ 
\hline
6  & \code{NamespaceFilter} & Sets and reverts a new namespace scope.\\ 
\hline
7  & \code{TypedefFilter} & Adds a type alias to the current scope via the 
                            C++ \code{typedef} statement.\\ 
\hline
8  & \code{UsingFilter} & Removes scope from a type by adding an alias in the 
                          current scope via the C++ \code{using} statement.\\ 
\hline
9  & \code{LinemarkerFilter} & Interprets \code{cpp} linemarker directives in order
                               to produce more useful debugging information in 
                               \cycpp.\\ 
\hline
10 & \code{NoteDecorationFilter} & Implements the \cycpp \code{#pragma cyclus note <dict>}
                                   directive by evaluating the contents of 
                                   \code{<dict>} and adding them to the archetype 
                                   annotations.\\ 
\hline
11 & \code{VarDecorationFilter} & Implements the \cycpp \code{#pragma cyclus var <dict>}
                                  directive for state variable annotations
                                  by evaluating the contents of \code{<dict>} in the current 
                                  context and queuing  them for the 
                                  next state variable declaration.\\ 
\hline
12 & \code{VarDeclarationFilter} & State variable declaration. Applies the results 
                                   of the immediately prior \code{VarDecorationFilter}
                                   as the state variable annotations. Furthermore, 
                                   this filter parses out the name of the 
                                   state variable, its index with respect to other 
                                   state variables on this class, and resolves its
                                   C++ type into an unambiguous form.\\ 
\hline
13 & \code{PragmaCyclusErrorFilter} & Throws errors if \code{#pragma cyclus} 
                                      directive is incorrectly implemented.
                                      This moves errors from happening at compile 
                                      or run time to \cycpp.\\
\hline
\end{tabular}
\label{pass2-filters}
\end{table}

In \cycpp, the only relevant type information is the name of the type. 
The concrete size in bits
of a type and the operations that are available for that type are not directly 
relevant. This is because the primary purpose of the type of a state variable is
to be able to fill in the appropriate values in the third pass. 

The canonical form of a type has the following spelling rules:
\begin{itemize}
    \item Primitive types (\code{int}, \code{double}, \code{std::string}, etc.) 
          and classes (\code{cyclus::Blob}, etc.) are spelled with strings 
          of the names.
    \item Template types (\code{std::vector}, \code{std::map}, etc.) are spelled 
          with lists of length of the number of template parameters plus one.
          The first element of the list is a string that represents the 
          template type (e.g. \code{std::pair}). The remaining elements of 
          the list represent the template parameter types, in order, and may 
          be either strings or lists.  For example, the type 
          \code{std::map<int, std::vector<double>>} would have the canonical form 
          of \code{['std::map', 'int', ['vector', 'double']]}.
    \item All namespaces must be included in the type name.
    \item Pointer and reference types are not allowed because these may not be 
          represented in the database.
\end{itemize}
The above rules create an accurate and language-independent
mechanism for spelling C++ types, including templates. The preprocessor is aware
of the following types that may be present in a \cyclus database in various 
combinations:
\begin{itemize}
    \item \textbf{Primitives:} \code{bool}, \code{int}, \code{float}, \code{double}, 
                               \code{std::string}
    \item \textbf{Known Classes:} \code{cyclus::Blob}, \code{boost::uuids::uuid}, 
                                  \code{cyclus::toolkit::ResBuff}, 
                                  \code{cyclus::toolkit::ResMap} 
    \item \textbf{Templates:} \code{std::vector}, \code{std::set}, \code{std::list}, 
                              \code{std::pair}, \code{std::map}
\end{itemize}

Resolving a canonical type name is necessarily a recursive process.
This is because aliases may point to other aliases --- not just primitive type names.
Thus to resolve an alias, one must walk through an arbitrarily deep graph of aliases 
to find the associated primitive.  For example, given that \code{myfloat} points 
to \code{float} (the primitive) and \code{mynumber} points to \code{myfloat}, 
if a state variable was declared as \code{myfloat} only one alias lookup would
be required whereas two would be required
if it was declared as \code{mynumber}. The canonical
form for all \code{float}, \code{myfloat}, and \code{mynumber} would all be 
\code{float}.  This is what \cycpp should record.  Templates must recursively 
determine the template type name and the types of all template parameters.
The canonical form of a type must be automatically computed to avoid an
entire class of typographic errors by the archetype developers.

Aside from the type system semantics, pass two represents a relatively straightforward
process of building up archetype information for later use. This
later use occurs during code generation in pass three of \cycpp. Conceptually, 
pass three is a more complex process than pass two because it must implement 
all of the member functions in Listing \ref{req-api}. In practice however, the body 
of each of these member functions follows its own pattern with respect to the 
state variables. Therefore, implementing pass three is significantly easier.

Much like pass two, pass three is also a state machine. The class that implements it is 
called \code{CodeGenerator} and the filters within this class implement 
the corresponding code generation routines.  While the \code{CodeGenerator}  does 
reuse some meta-data accumulation filters, it largely relies on the results of 
pass two for archetype and state variable information.  The only data that cannot be 
reused and is recomputed is that which pertains to the scope of each line of C++ code.

Pass three traverses all lines of the source code for a third time.
On this pass, the \code{CodeGenerator} will replace certain \code{#pragma cyclus}
directives with the generated implementations.  Pass three may act on the output of
\code{cpp}, the results of pass one.  However, it is more common for this 
to act on the original source and header files.  This requires that the 
archetype developer write in mostly normative C++ and not abuse the C
preprocessor. They must 
avoid double include errors and other downstream issues with 
compilation. The results of pass three, therefore, are a new version of the 
archetype
source code that differs only in that it contains automatically implemented 
member functions.

Table \ref{pass3-filters} displays the filters that the \code{CodeGenerator} 
employs in order of precedence. These filters overlap somewhat with
those of the \code{StateAccumulator}. This enables the efficient reuse
of filters between state machines.

\begin{table}
\caption{\cyclus Preprocessor Pass 3 Filters, higher order filters have 
         lower execution precedence.}
\begin{tabular}[htb]{|p{0.05\linewidth}|p{0.33\linewidth}|p{0.6\linewidth}|}
\hline
\textbf{order} & \textbf{filter} & \textbf{description} \\
\hline
1  & \code{InitFromCopyFilter} & Implements code generation for copy-constructor-like 
                                 \code{InitFrom()} member function. This may be called
                                 with the 
                        \code{#pragma cyclus [def\|decl\|impl] initfromcopy [classname]}
                                 directive.\\
\hline
2  & \code{InitFromDbFilter} & Implements code generation for database constructor 
                               \code{InitFrom()} member function. This may be called
                               with the 
                        \code{#pragma cyclus [def\|decl\|impl] initfromdb [classname]}
                               directive.\\
\hline
3  & \code{InfileToDbFilter} & Implements code generation for the \code{InfileToDb()} 
                               member function that converts an input file into its 
                               database representation. This may be called with the 
                        \code{#pragma cyclus [def\|decl\|impl] infiletodb [classname]}
                               directive.\\
\hline
4  & \code{CloneFilter} & Implements code generation for the \code{Clone()} member
                          function that clones prototypes. This may be called
                          with the 
                        \code{#pragma cyclus [def\|decl\|impl] clone [classname]}
                          directive.\\
\hline
5  & \code{SchemaFilter} & Implements code generation for the \code{schema()} member
                           function that returns the \gls{RNG} schema of the archetype
                           for input file validation. This may be called with the 
                        \code{#pragma cyclus [def\|decl\|impl] schema [classname]}
                           directive.\\
\hline
6  & \code{AnnotationsFilter} & Implements code generation for the \code{annotations()} 
                                member function that returns the archetype metadata
                                that was compiled during \cycpp pass two.
                                This may be called with the 
                        \code{#pragma cyclus [def\|decl\|impl] annotations [classname]}
                                directive.\\
\hline
7  & \code{InitInvFilter} & Implements code generation for the \code{InitInv()} 
                            member function that sets the initial resource inventories 
                            of the agent. This may be called with the 
                        \code{#pragma cyclus [def\|decl\|impl] initinv [classname]}
                            directive.\\
\hline
8  & \code{SnapshotInvFilter} & Implements code generation for the \code{SnapshotInv()} 
                                member function that writes inventories to the 
                                database. This may be called with the 
                        \code{#pragma cyclus [def\|decl\|impl] snapshotinv [classname]}
                                directive.\\
\hline
9  & \code{SnapshotFilter} & Implements code generation for the \code{Snapshot()} 
                             member function that writes state variables to the
                             database. This may be called with the 
                        \code{#pragma cyclus [def\|decl\|impl] snapshot [classname]}
                             directive.\\
\hline
10 & \code{ClassFilter} & Sets the current class name and scope.\\
\hline
11 & \code{AccessFilter} & Sets the current access control level, either 
                           \code{public}, \code{private}, or \code{protected}.\\
\hline
12 & \code{NamespaceAliasFilter} & Implements namespace aliasing in the current 
                                   scope.\\ 
\hline
13 & \code{NamespaceFilter} & Sets and reverts a new namepsace scope.\\ 
\hline
\end{tabular}
\label{pass3-filters}
\end{table}

\begin{table}
\caption{\cyclus Preprocessor Pass 3 Filters (part 2).}
\begin{tabular}[htb]{|p{0.05\linewidth}|p{0.33\linewidth}|p{0.6\linewidth}|}
\hline
\textbf{order} & \textbf{filter} & \textbf{description} \\
\hline
14 & \code{VarDecorationFilter} & Implements the \cycpp \code{#pragma cyclus var <dict>}
                                  directive for state variable annotations 
                                  by evaluating the contents of \code{<dict>} in the current 
                                  context and queuing them for the 
                                  next state variable declaration.\\ 
\hline
15 & \code{VarDeclarationFilter} & State variable declaration. Applies the results 
                                   of the immediately prior \code{VarDecorationFilter}
                                   as the state variable annotations. Furthermore, 
                                   this filter parses out the name of the 
                                   state variable, its index with respect to other 
                                   state variables on this class, and resolves its
                                   C++ type into an unambiguous form.\\ 
\hline
16 & \code{LinemarkerFilter} & Interprets \code{cpp} linemarker directives in order
                               to produce more useful debugging information in 
                               \cycpp.\\ 
\hline
17 & \code{DefaultPragmaFilter} & Implements the default code generation directive,
                                  \code{#pragma cyclus [def\|decl\|impl]}. This 
                                  calls the other code generation
                                  filters to obtain member function implementations.\\
\hline
18 & \code{PragmaCyclusErrorFilter} & Throws errors if \code{#pragma cyclus} 
                                      directive is incorrectly implemented.
                                      This moves errors from happening at compile 
                                      or run time to \cycpp.\\
\hline
\end{tabular}
\label{pass3-filters-2}
\end{table}

When all of the pieces of \cycpp are brought together, the benefits scale as the 
number of state variables times the number of code generated member functions 
(currently nine). This implies roughly an order of magnitude savings on the number of 
lines that an archetype developer must write per state variable. Moreover, 
exponential savings come from the fact that the archetype developers do not need
to understand the details of the \cyclus interface.
Even for a developer who knows the \cyclus
interface completely, there is still a factor of ten less code to write. Consider 
again the simple \code{Reactor} example presented in Listing \ref{rx-eg}.  These twelve 
lines of code are transformed into 112 lines by \cycpp, the results of which
are shown in Listing \ref{rx-eg-cycpp}. 

\begin{lstlisting}[caption={Simple Reactor Archetype After Preprocessing with \cycpp, 
                            line marker directives have been removed for space}, 
                   label=rx-eg-cycpp]
class Reactor : public cyclus::Facility {
 public:
  Reactor (cyclus::Context* ctx) {};
  virtual ~Reactor() {};

  virtual void InitFrom(Reactor* m) {
    flux = m->flux;
    power = m->power;
    shutdown = m->shutdown;
  };

  virtual void InitFrom(cyclus::QueryableBackend* b) {
    cyclus::QueryResult qr = b->Query("Info", NULL);
    flux = qr.GetVal<double>("flux");
    power = qr.GetVal<float>("power");
    shutdown = qr.GetVal<bool>("shutdown");
  };

  virtual void InfileToDb(cyclus::InfileTree* tree, cyclus::DbInit di) {
    tree = tree->SubTree("config/*");
    cyclus::InfileTree* sub;
    int i;
    int n;
    flux = cyclus::OptionalQuery<double>(tree, "flux", 4e+14);
    power = cyclus::OptionalQuery<float>(tree, "power", 1000);
    shutdown = cyclus::Query<bool>(tree, "shutdown");
    di.NewDatum("Info")
    ->AddVal("flux", flux)
    ->AddVal("power", power)
    ->AddVal("shutdown", shutdown)
    ->Record();
  };

  virtual cyclus::Agent* Clone() {
    Reactor* m = new Reactor(context());
    m->InitFrom(this);
    return m;
  };

  virtual std::string schema() {
    return ""
      "<interleave>\n"
      "<optional>\n"
      "    <element name=\"flux\">\n"
      "        <data type=\"double\" />\n"
      "    </element>\n"
      "</optional>\n"
      "<optional>\n"
      "    <element name=\"power\">\n"
      "        <data type=\"float\" />\n"
      "    </element>\n"
      "</optional>\n"
      "<element name=\"shutdown\">\n"
      "    <data type=\"boolean\" />\n"
      "</element>\n"
      "</interleave>\n"
      ;
  };

  virtual Json::Value annotations() {
    Json::Value root;
    Json::Reader reader;
    bool parsed_ok = reader.parse(
      "{\"name\":\"Reactor\",\"entity\":\"unknown\",\"parents\":[],"
      "\"all_parents\":[],\"vars\":{\"flux\":{\"default\":4000000"
      "00000000.0,\"units\":\"n/cm2/2\",\"type\":\"double\",\"inde"
      "x\":0},\"power\":{\"default\":1000,\"units\":\"MWe\",\"type\""
      ":\"float\",\"index\":1},\"shutdown\":{\"doc\":\"Are we "
      "operating?\",\"type\":\"bool\",\"index\":2}}}", root);
    if (!parsed_ok) {
      throw cyclus::ValueError("failed to parse annotations for Reactor.");
    }
    return root;
  };

  virtual void InitInv(cyclus::Inventories& inv) {
  };

  virtual cyclus::Inventories SnapshotInv() {
    cyclus::Inventories invs;
    return invs;
  };

  virtual void Snapshot(cyclus::DbInit di) {
    di.NewDatum("Info")
    ->AddVal("flux", flux)
    ->AddVal("power", power)
    ->AddVal("shutdown", shutdown)
    ->Record();
  };

 private:
  #pragma cyclus var {'default': 4e14, 'units': 'n/cm2/2'}
  double flux;

  #pragma cyclus var {'default': 1000, 'units': 'MWe'}
  float power;

  #pragma cyclus var {'doc': 'Are we operating?'}
  bool shutdown;
};
\end{lstlisting}

Of course, archetypes may be much more complex than the \code{Reactor} example.
This archetype does not participate in resource exchange, take advantage of 
the reflection features, use available annotations, or have more than 
a handful of state variables.  Yet, even here, the value of a code generating
preprocessor is readily apparent.

\subsection{Database Backends \& Types}

\Cyclus transparently supports a potentially limitless number of different database 
backends. Currently, two reference backends exist: \gls{SQLite} \cite{owens2006definitive} 
and \gls{HDF5} \cite{folk2011overview}. These two represent relational and hierarchical 
databases, respectively, and have different underlying design philosophies.
Future formats that could be supported include plain text 
\gls{CSV} files or \gls{JSON}.  Given the wide range of potential uses cases, \cyclus must be able 
to execute based on only the feature set that is common among these formats.
Features that are not available in a single format must either be provided by \cyclus 
itself or become optional in the backend interface.

The type system and reflection provided by \cycpp 
allow archetypes to represent themselves in the database. Though this 
reflection can be taken advantage of and used elsewhere, it was initially
implemented because of the need to restart a simulation.
The database model extends well beyond the needs of the archetypes alone by 
serving as the fundamental on-disk representation for all \cyclus input and output.

\Cyclus databases follow these fundamental abstractions:
\begin{itemize}
    \item All data are stored in tables with named columns,
    \item Tables live in a flat hierarchy,
    \item Columns may have any type described by the \cyclus type system, and
    \item All tables must have a \code{SimId} column which uniquely and 
          universally identifies the simulation.
\end{itemize}
Therefore, common databases notions such as the shape of a column, raw arrays, or
queryability must be optional, omitted, or implemented inside the backend itself.
Queryability is the most important, and is available in both \gls{HDF5} and 
\gls{SQLite}. If a database format or its backend lacks 
queryability (such as \gls{CSV}), then it is impossible to start or restart a 
\cyclus simulation from it. Though, such a format may be used in 
conjunction with another, queryable format. 

The database backends are deeply tied to the \cyclus type system. Types in \cyclus
are represented by unique integers that map to C++ types by the \code{enum} called \code{DbTypes}.
Simple types are represented by simple names: \code{float} becomes
\code{FLOAT}, which is assigned to the identifier 2. Container types, such as \code{vector},
are more complex in that each template specification (\code{vector<int>}) has
its own type in the \cyclus type system.  Containers are further delineated
as either fixed-length or \gls{VL}. Thus even the 
relatively simple C++ type \code{std::vector<int>} receives two entries in the 
\cyclus type system: \code{VECTOR_INT} and \code{VL_VECTOR_INT}, which are given 
the identifiers 10 and 11. 
Thus the number of types in the \cyclus type system is obtained as 
2 to the power of the \emph{rank} or the total number of variable length parameters, 
including nettings.
For example, the number of \cyclus types for \code{std::map<int, std::string>} is 
four (\code{MAP_INT_STRING}, \code{VL_MAP_INT_STRING}, \code{MAP_INT_VL_STRING}, 
\code{VL_MAP_INT_VL_STRING}) since both maps and strings may be variable length.
Table \ref{some-types} displays a sampling of types currently implemented in 
\cyclus.
Each backend determines which types it wishes to support, though this must extend 
to a relatively robust subset in order to run even simple simulations.

\begin{table}
\caption{Sample \cyclus Types}
\centering
\begin{tabular}[htb]{|c|l|l|c|}
\hline
\textbf{id} & \textbf{name} & \textbf{C++ type} & \textbf{rank} \\
\hline
0 & \code{BOOL} & \code{bool} & 0 \\
1 & \code{INT} & \code{int} & 0 \\
2 & \code{FLOAT} & \code{float} & 0 \\
3 & \code{DOUBLE} & \code{double} & 0 \\
4 & \code{STRING} & \code{std::string} & 1 \\
5 & \code{VL_STRING} & \code{std::string} & 1 \\
6 & \code{BLOB} & \code{cyclus::Blob} & 0 \\
7 & \code{UUID} & \code{boost::uuids::uuid} & 0 \\
8 & \code{VECTOR_BOOL} & \code{std::vector<bool>} & 1 \\
9 & \code{VL_VECTOR_BOOL} & \code{std::vector<bool>} & 1 \\
10 & \code{VECTOR_INT} & \code{std::vector<int>} & 1 \\
11 & \code{VL_VECTOR_INT} & \code{std::vector<int>} & 1 \\
 & $\cdots$ & $\cdots$ & \\
32 & \code{SET_STRING} & \code{std::set<std::string>} & 2 \\
33 & \code{VL_SET_STRING} & \code{std::set<std::string>} & 2 \\
34 & \code{SET_VL_STRING} & \code{std::set<std::string>} & 2 \\
35 & \code{VL_SET_VL_STRING} & \code{std::set<std::string>} & 2 \\
 & $\cdots$ & $\cdots$ & \\
42 & \code{LIST_INT} & \code{std::list<int>} & 1 \\
43 & \code{VL_LIST_INT} & \code{std::list<int>} & 1 \\
 & $\cdots$ & $\cdots$ & \\
57 & \code{PAIR_INT_INT} & \code{std::pair<int, int>} & 0 \\
 & $\cdots$ & $\cdots$ & \\
104 & \code{MAP_STRING_STRING} & \code{std::map<std::string, std::string>} & 3 \\
105 & \code{VL_MAP_STRING_STRING} & \code{std::map<std::string, std::string>} & 3 \\
106 & \code{MAP_STRING_VL_STRING} & \code{std::map<std::string, std::string>} & 3 \\
107 & \code{VL_MAP_STRING_VL_STRING} & \code{std::map<std::string, std::string>} & 3 \\
 & $\cdots$ & $\cdots$ & \\
120 & \code{MAP_VL_STRING_STRING} & \code{std::map<std::string, std::string>} & 3 \\
121 & \code{VL_MAP_VL_STRING_STRING} & \code{std::map<std::string, std::string>} & 3 \\
122 & \code{MAP_VL_STRING_VL_STRING} & \code{std::map<std::string, std::string>} & 3 \\
123 & \code{VL_MAP_VL_STRING_VL_STRING} & \code{std::map<std::string, std::string>} & 3 \\
 & $\cdots$ & $\cdots$ & \\
\hline
\end{tabular}
\label{some-types}
\end{table}

A key database-enabling feature of the \cyclus type system is that 
the values of all types must be directly \emph{hashable} using a cryptographic hash.
The directness again implies that pointer and reference types are not allowed. For most 
database backends, indirection is not a supported feature. For those backends which 
do support indirection, such as \gls{HDF5} and its linking mechanism, there is not a 
clear translation from indirection in memory to indirection on disk.   
Requiring hashable data types avoids several classes of errors entirely.  

Hashability serves a dual role with respect to the backends. The first is that 
it provides a mechanism for uniquely identifying all elements of a type within 
reason. \Cyclus uses the standard \gls{SHA1} \cite{eastlake2001us} algorithm to compute 
hash values as 160-bit integers.  Thus for types with a fixed bit width less than 
160, such as \code{int} (typically 32-bits) or \code{double} (64-bits), every 
element is uniquely identifiable. For variable-length data types or very long types,
the probability of a hash collision is only $2^{-160}$, which is approximately 
equal to $10^{-48}$.  This is an astronomically small possibility, even over the 
course of billions of simulations.  Thus, backends may use the hash to automatically
de-duplicate data and store every unique value only once.

The second purpose for hashing is to allow backends to implement the storage 
of variable-length types as a \emph{bidirectional hash map} using the 
\gls{SHA1} as a key.  This data structure is an associative array in which the values are 
uniquely determined from the keys \emph{and} the keys are uniquely
determined from 
the values. Furthermore, in \cyclus, the keys of this data structure are simply the 
hashes themselves. This differs from a typical hash map (e.g. Python dictionaries) 
in that they only require that values may be determined from the keys and only the
keys must be unique.  With a bidirectional hash map, knowing either the key or the value
will provide the value or the key, respectively.  The \gls{HDF5} backend takes advantage 
of this data structure to store variable-length data in a 5-dimensional sparse array.
The hash is chopped up into an array of five 32-bit unsigned integers that 
index into this sparse array. Then the hash is stored in the table and used to access 
the value in the corresponding sparse array for that type.  This creates an efficient 
mechanism for storing vast amounts of potentially redundant variable-length data in 
a manner that mirrors the column storage for primitive types (\code{bool}, \code{int}, 
etc.). Since the hash is itself the index into a sparse array, the overhead from 
this lookup is minimal as compared with other parts of backend infrastructure.

Archetype developers may create their own custom tables in the database as well.
This is done through using the backend interface directly in the archetype. 
Data that are fully dependent parameters of the archetype are not appropriate 
as state variables and thus should not be stored in this way.  Custom tables 
have the same restrictions as other parts of the database as well as 
the additional restriction that they cannot reuse the table names that 
\cyclus itself uses. Writing to such tables is reserved for
the kernel. 
Table \ref{std-tabs} shows the standard tables generated by \cyclus.
Distinct from the previously mentioned custom tables, the kernel will also produce 
tables whose names are based on the 
archetype specification for representing the archetype on disk.  These tables
are also reserved for the kernel alone. 

\begin{table}
\caption{Standard Tables Reserved by the \Cyclus Kernel, Columns given in order with 
names and types.}
\centering
\begin{tabular}[htb]{|llp{0.75\linewidth}|}
\hline
\textbf{name} & \textbf{type} & \textbf{description} \\
\hline
\multicolumn{3}{|p{0.95\linewidth}|}{\textbf{Resources Table:} 
Encodes a heritage tree for all resources. 
Because resources 
are tracked as immutable objects, every time a resource changes in the 
simulation (split, combined, transmuted, decayed, etc.), a new entry is added
to this table. If two resources are 
combined, then the new resource entry will have the identifiers of the 
other two in its ``Parent1'' and ``Parent2'' columns. 
The \code{Resources}
table does not encode any information about where a resource exists. This information 
can be inferred from the \code{ResCreators} and 
\code{Transactions} tables.}\\
& & \\
\code{SimId} & \code{UUID} & Simulation identifier \\
\code{ResourceId} & \code{INT} & The unique ID for this resource entry. \\
\code{ObjId} & \code{INT} & A resources object id (\code{obj_id}) as it existed 
                            during the simulation simulation.\\
\code{Type} & \code{VL_STRING} & One of ``Material'' or ``Product''. These two types 
                                 of resources have different internal state stored 
                                 in different tables. If the type is product, 
                                 then the internal state can be found in the 
                                 \code{Products} table. If it is material, 
                                 then it is in the \code{Compositions} table.\\
\code{TimeCreated} & \code{INT} & The simulation time step at which this resource 
                                  state came into existence.\\
\code{Quantity} & \code{DOUBLE} & Amount of the resource in ``kg'' for material 
                                  resources. Amount in terms of the specific quality 
                                  for product resources.\\
\code{Units} & \code{VL_STRING} & ``kg'' for all material resources, ``NONE'' for 
                                   product resources.\\
\code{QualId} & \code{INT} & Used to identify the corresponding internal-state 
                             entry (or entries) in the \code{Products} or 
                             \code{Compositions} table depending on the resource type.\\
\code{Parent1} & \code{INT} & If a resource was newly created, this is zero. If this 
                              resource came from transmutation, 
                              combining, splitting, or decay then this is the 
                              parent ResourceId.\\
\code{Parent2} & \code{INT} & If a resource was newly created, this is zero. If this 
                              resource came from transmutation, decay, 
                              or splitting, this is also zero. If the resource 
                              came from combining then this is the second 
                              parent's ResourceId.\\
\hline
\multicolumn{3}{|p{0.95\linewidth}|}{\textbf{Compositions Table:}
Stores compositions. A composition consists of one or more nuclides and their respective mass fractions. 
Each nuclide within a composition has its own row but the same \code{QualId}.}\\
& & \\
\code{SimId} & \code{UUID} & Simulation identifier \\
\code{QualId} & \code{INT} & Key to associate this composition with one or more 
                             entries in the \code{Resources} table.\\
\code{NucId} & \code{INT} & Nuclide identifier in zzzaaammmm form.\\
\code{MassFrac} & \code{DOUBLE} & Mass fraction for the nuclide in this composition.\\
\hline
\multicolumn{3}{|p{0.95\linewidth}|}{\textbf{Recipes Table:} Stores composition names.}\\
& & \\
\code{SimId} & \code{UUID} & Simulation identifier \\
\code{Recipe} & \code{VL_STRING} & Recipe name as given in the input file.\\
\code{QualId} & \code{INT} & Key to identify the composition for this recipe in the 
                             \code{Compositions} table.\\
\hline
\end{tabular}
\label{std-tabs}
\end{table}

\begin{table}
\caption{Standard Tables Reserved by the \Cyclus Kernel (cont.)}
\centering
\begin{tabular}[htb]{|llp{0.75\linewidth}|}
\hline
\multicolumn{3}{|p{0.95\linewidth}|}{\textbf{Products Table:} Stores
product information regarding quality.}\\
& & \\
\code{SimId} & \code{UUID} & Simulation identifier \\
\code{QualId} & \code{INT} & Key to associate this quality with one or more entries 
                             in the \code{Resources} table.\\
\code{Quality} & \code{VL_STRING} & Describes a product's quality (e.g. ``bananas'', 
                                    ``KWh'', etc.).\\
\hline
\multicolumn{3}{|p{0.95\linewidth}|}{\textbf{ResCreators Table:} Stores
instances of a new resource being created by an agent.}\\
& & \\
\code{SimId} & \code{UUID} & Simulation identifier. \\
\code{ResourceId} & \code{INT} & ID of a resource that was created during 
                                 the simulation.\\
\code{AgentId} & \code{INT} & ID of the agent that created the resource associated 
                              with the \code{ResourceId}.\\
\hline
\multicolumn{3}{|p{0.95\linewidth}|}{\textbf{AgentEntry Table:} Stores
a row of information for each agent that enters the simulation.}\\
& & \\
\code{SimId} & \code{UUID} & Simulation identifier. \\
\code{AgentId} & \code{INT} & Every agent in a simulation gets its own, unique ID.\\
\code{Kind} & \code{VL_STRING} & Entity type. One of ``Region'', ``Inst'', ``Facility'', 
                                 or ``Agent''.\\
\code{Spec} & \code{VL_STRING} & The single-string of the agent specification.\\
\code{Prototype} & \code{VL_STRING} & The prototype name, as defined in the input file, 
                                      that was used to create this agent.\\
\code{ParentId} & \code{INT} & The \code{AgentId} of the parent agent that 
                               built or created this agent.\\
\code{Lifetime} & \code{INT} & Number of time steps an agent is designed to operate 
                               over. -1 indicates an infinite lifetime.\\
\code{EnterTime} & \code{INT} & The time step when the agent entered 
                                the simulation.\\
\hline
\multicolumn{3}{|p{0.95\linewidth}|}{\textbf{AgentExit Table:} Stores
a row of information for agents that leave a simulation. 
If this table does not exist, then no agents were decommissioned in the simulation.}\\
& & \\
\code{SimId} & \code{UUID} & Simulation identifier. \\
\code{AgentId} & \code{INT} & Key to the \code{AgentId} on the \code{AgentEntry} table.\\
\code{ExitTime} & \code{INT} & The time step when the agent  
                               exited the simulation.\\
\hline
\multicolumn{3}{|p{0.95\linewidth}|}{\textbf{Transactions Table:} Every single resource 
                                     transfer between two agents is recorded as a row 
                                     in this table.}\\
& & \\
\code{SimId} & \code{UUID} & Simulation identifier. \\
\code{TransactionId} & \code{INT} & A unique identifier for this resource transfer.\\
\code{SenderId} & \code{INT} & \code{AgentId} for the sending agent.\\
\code{ReceiverId} & \code{INT} & \code{AgentId} for the receiving agent.\\
\code{ResourceId} & \code{INT} & Key to the entry in the \code{Resources} table that 
                                 describes the transferred resource.\\
\code{Commodity} & \code{VL_STRING} & The commodity under which this transfer was 
                                      negotiated.\\
\code{Time} & \code{INT} & The time step at which the resource transfer took place.\\
\end{tabular}
\label{std-tabs-2}
\end{table}

\begin{table}
\caption{Standard Tables Reserved by the \Cyclus Kernel (cont.)} 
\centering
\begin{tabular}[htb]{|llp{0.65\linewidth}|}
\hline
\multicolumn{3}{|p{0.95\linewidth}|}{\textbf{Info Table:} Stores a row of
information for each simulation that describes global simulation parameters 
and \Cyclus dependency version information.}\\
& & \\
\code{SimId} & \code{UUID} & Simulation identifier. \\
\code{Handle} & \code{VL_STRING} & A user-specified value from the input 
                                   file allowing for human identification of 
                                   simulations in a database.\\
\code{InitialYear} & \code{INT} & The year in which time step zero occurs.\\
\code{InitialMonth} & \code{INT} & The month that time step zero represents.\\
\code{Duration} & \code{INT} & The length of the simulation in time steps.\\
\code{ParentSimId} & \code{UUID} &  The SimId for this simulation's parent. Zero if 
                                    this simulation has no parent.\\
\code{ParentType} & \code{VL_STRING} &  One of:
    \begin{enumerate}
        \item ``init'' for simulations that are not based on any other simulation.
        \item ``restart'' for simulations that were restarted another simulation's 
              snapshot.
        \item ``branch'' for simulations that were started from a perturbed state of 
              another simulation's snapshot.
    \end{enumerate}\\
\code{BranchTime} & \code{INT} & Zero if this was not a restarted or branched 
                                 simulation. Otherwise, the time step of the 
                                 \code{ParentSim} at which the restart or branch 
                                 occurred.\\
\code{CyclusVersion} & \code{VL_STRING} & Version of \Cyclus used to run this 
                                          simulation.\\
\code{CyclusVersionDescribe} & \code{VL_STRING} & Detailed \Cyclus version info 
                                                  (with commit hash).\\
\code{SqliteVersion} & \code{VL_STRING} & SQLite version information.\\
\code{Hdf5Version} & \code{VL_STRING} & HDF5 version information.\\
\code{BoostVersion} & \code{VL_STRING} & Boost version information.\\
\code{LibXML2Version} & \code{VL_STRING} & libxml2 version information.\\
\code{CoinCBCVersion} & \code{VL_STRING} & COIN version information.\\
\hline
\end{tabular}
\label{std-tabs-3}
\end{table}

\begin{table}
\caption{Standard Tables Reserved by the \Cyclus Kernel (cont.)} 
\centering
\begin{tabular}[htb]{|llp{0.75\linewidth}|}
\hline
\multicolumn{3}{|p{0.95\linewidth}|}{\textbf{Finish Table:} Stores one row of
information for each simulation.}\\
& & \\
\code{SimId} & \code{UUID} & Simulation identifier. \\
\code{EarlyTerm} & \code{BOOL} & True if the simulation terminated early and did 
                                 not complete normally. False otherwise.\\
\code{EndTime} & \code{INT} & The time step at which the simulation ended.\\
\hline
\multicolumn{3}{|p{0.95\linewidth}|}{\textbf{InputFiles Table:} Stores the 
simulation input.}\\
& & \\
\code{SimId} & \code{UUID} & Simulation identifier. \\
\code{Data} & \code{BLOB} & A dump of the entire input file used for this simulation.\\
\hline
\multicolumn{3}{|p{0.95\linewidth}|}{\textbf{DecomSchedule Table:} Stores
information regarding when agents are scheduled to be decommissioned in the
simulation. If a simulation ended before time reached the scheduled time, the 
agent would not have been decommissioned, but this table still includes the 
schedule information.}\\
& & \\
\code{SimId} & \code{UUID} & Simulation identifier. \\
\code{AgentId} & \code{INT} & ID of the agent that is to be decommissioned.\\
\code{SchedTime} & \code{INT} & The time step on which this decommissioning event was
                                created.\\
\code{DecomTime} & \code{INT} & The time step on which the agent was (or would have
                                been) decommissioned.\\
\hline
\multicolumn{3}{|p{0.95\linewidth}|}{\textbf{BuildSchedule Table:} Stores
information regarding when agents are scheduled to be built in the simulation.  
If a simulation ended before time reached the scheduled time, the agent would 
not have been built, but this table still includes the schedule information.} \\
& & \\
\code{SimId} & \code{UUID} & Simulation identifier. \\
\code{ParentId} & \code{INT} & The ID of the agent that will become this new agent's
                               parent.\\
\code{Prototype} & \code{VL_STRING} & The name of the agent prototype that will be 
                                      used to generate the new agent. This corresponds 
                                      to the prototypes defined in an input files.\\
\code{SchedTime} & \code{INT} & The time step on which this build event was created.\\
\code{BuildTime} & \code{INT} & The time step on which the agent was (or would have
                                been) built and deployed into the simulation.\\
\hline
\multicolumn{3}{|p{0.95\linewidth}|}{\textbf{Snapshots Table:} Stores
entries containing information about every snapshot made during the
simulation. All times in this table are candidates for 
simulation restart and branching.}\\
& & \\
\code{SimId} & \code{UUID} & Simulation identifier. \\
\code{Time} & \code{INT} & The time step a snapshot was taken for this simulation.\\
\hline
\end{tabular}
\label{std-tabs-4}
\end{table}

Though the database backend implementation and the associated type system may be 
complex to implement, its usage is mostly hidden to users and 
archetype developers through code generation. Even for complex template types, 
the \cyclus type system
allows archetypes to be as expressive as needed to fit the model. 

\subsection{JSON Annotations}

Archetype metadata annotation is an important part of \cyclus because it allows for 
reflection on the archetype classes. Well-defined metadata entries are 
described in Tables \ref{sv-anno} \& \ref{ag-anno}. 
Additionally, archetype developers may supply 
any other information and ascribe to it the semantics that they desire.
This is done simply by adding undefined keys to the \code{#pragma cyclus var} and
\code{#pragma cyclus note} \cycpp directives.
While this ensures that the metadata is robust to future changes and archetype developer
customization, the annotations have implications for the \cyclus and archetype 
implementations.

Allowing for unknown metadata keys with unknown types for each value implies that 
the metadata is \emph{unstructured} \cite{feldman2007text}. From a C++ implementation 
standpoint, this means that there is no class or struct that can be declared whose
member variables encompass all possible metadata without at least one of those
members being a pointer or reference. This is because determining the type 
of a blob of memory at runtime in C++ must use \code{void*}, \code{char*}, 
or other pointer indirection. Representing annotation in memory is 
more complex than in a single metadata class.

\acrlong{JSON} and its derivatives are largely acknowledged as sufficiently expressive 
formats for unstructured data \cite{moniruzzaman2013nosql}. This is because the
\gls{JSON} primitives, which include integers, floats, strings, booleans, null, arrays, 
and objects (hash tables with string keys), are easily translatable into native
data structures in most modern programming languages such as Python and C++. 
Furthermore, the \gls{JSON} syntax is concise and intuitive. \gls{XML} could have been used 
as an alternative format, but a schema and the translation to native data structures 
would have to be handled manually. \gls{YAML} \cite{ben2009yaml} or Python itself offer 
more likely  alternatives to \gls{JSON} but require a more sophisticated interpreters 
corresponding to their more powerful syntax. 

\gls{JSON} is a metadata representation that resides on disk, 
not in memory. Translation from plain text \gls{JSON} to C++ data structures is 
performed via the JsonCpp software \cite{eltuhamy2014native}. This does the work 
of parsing \gls{JSON} code, implementing the \gls{JSON} type system, and translating back and
forth between \gls{JSON} types and C++ types. It provides a fully 
introspective 
container for arbitrary metadata called \code{Json::Value}. An instance of this 
class is precisely what the \code{annotations()} archetype member function returns.
Therefore, the entire metadata workflow is as follows:
\begin{enumerate}
    \item Archetype developer writes metadata using annotation directives as a
          Python dictionary with string keys.
    \item \cycpp parses, evaluates, and accumulates the annotations into a single 
          metadata dictionary per archetype during pass two of preprocessing.
    \item Pass three of \cycpp converts the metadata to a \gls{JSON}-formatted 
          string using 
          \gls{JSON} utilities in the Python standard library. This comprises
          the majority of the 
          code generated for \code{annotations()} member
          function.
    \item When \code{annotation()} is called from C++, the \gls{JSON} string is parsed 
          by JsonCpp and a new instance of \code{Json::Value} is returned that
          corresponds to the metadata.
\end{enumerate}
In this way, \gls{JSON} is used as an exchange format between Python and C++, between 
archetype developer and user, and between compile time and run time.

\subsection{XML Validation}

A key benefit to the \Cyclus simulation infrastructure is the runtime
guarantee
of valid input files provided to both users and archetype developers. Developers
are guaranteed valid construction of archetypes within a simulation, and users
are notified immediately if a given input file would have resulted in invalid
archetype construction. The processes required to provide such guarantees are
implemented using robust schema validation with \gls{XML} and
\gls{RNG} \cite{clark2001relax}.

For a given archetype, a developer defines the expected input structure in the
\code{schema()} function manually or via the \code{#pragma cyclus var}
preprocessor directive. Upon initiating a \Cyclus simulation (i.e., at run
time), a master schema is generated by combining the schema of all discovered
archetypes on a given computing system and inserting the collection into a
general \Cyclus schema. The master \Cyclus schema is used to define
simulation-level input as well as general entity input. For example, all
\code{cyclus::Facility} archetypes have input parameters common to the
\code{cyclus::Facility} entity, e.g., a name and a lifetime, and input
parameters specific to their archetype. Listing \ref{fac-schema} shows a
section of a generated \Cyclus master schema pertaining to the
\code{cyclus::Facility} entity input on a computing system on which archetypes
named \code{Reactor}, \code{Source}, and \code{Sink} were installed. Listing
\ref{rx-schema} shows the generated schema for the simple \code{Reactor}
archetype discussed in section \S \ref{subsec-ppgc}.

\lstset{language=XML}
\begin{lstlisting}[caption={Generated \Cyclus Facility Schema for a Computing System with 
      \code{Reactor}, \code{Source}, and \code{Sink} Archetypes Installed}, 
    label=fac-schema]
<oneOrMore>
  <element name="facility">
    <element name="name"> 
      <text/> 
    </element>
    <optional>
      <element name="lifetime"> 
        <data type="nonNegativeInteger"/> 
      </element>
    </optional>
    <element name="config">
      <choice>
        <ref name="Reactor"/ >
        <ref name="Source"/ >
        <ref name="Sink"/ >
      </choice>
    </element>
  </element>
</oneOrMore>
\end{lstlisting}

\lstset{language=XML}
\begin{lstlisting}[caption={Generated Simple Reactor Schema}, 
                   label=rx-schema]
<element name="Reactor">
  <interleave>
    <optional>
      <element name="flux">
        <data type="double" />
      </element>
    </optional>
    <optional>
      <element name="power">
        <data type="float" />
      </element>
    </optional>
    <element name="shutdown">
      <data type="boolean" />
    </element>
  </interleave>
</element>
\end{lstlisting}

After generating master \Cyclus schema, user input is provided to an instance
of the \code{cyclus::RelaxNGValidator} class. This class utilizes the C++ libxml2
library to validate the user input against the generated \gls{RNG} schema. 

If
input validation is successful, the defined simulation is then instantiated and
executed. A section of valid input for the \code{Reactor} is shown in Listing
\ref{valid-xml}. Note that because the \code{<optional>} tag is utilized in the
schema, not all parameters are required to be specified. Furthermore, default
values defined in the \code{Reactor} \code{#pragma cyclus var} annotations are
used for the unspecified parameters.

\lstset{language=XML}
\begin{lstlisting}[caption={A Valid Input Snippet for the Simple Reactor}, 
                   label=valid-xml]
<facility>
  <name>SomeReactor</name>
  <lifetime>600</lifetime>
  <config>
    <Reactor>
      <power>1150</power>
    </Reactor>
  </config>
</facility>
\end{lstlisting}

If the input is determined to be invalid, an error is raised without
beginning the simulation. Listing \ref{invalid-xml} shows an example of input for
the \code{Reactor} that is invalid because the \code{power} input parameter
type is not a \code{float}; \Cyclus fails immediately with the error message
shown in Listing \ref{lst-error}.

\lstset{language=XML}
\begin{lstlisting}[caption={An Invalid Input Snippet for the Simple Reactor}, 
                   label=invalid-xml]
<facility>
  <name>SomeReactor</name>
  <lifetime>600</lifetime>
  <config>
    <Reactor>
      <power>magic</power>
    </Reactor>
  </config>
</facility>
\end{lstlisting}

\lstset{language=bash}
\begin{lstlisting}[caption={A \Cyclus Error Message from the Invalid 
      Input in Listing \ref{invalid-xml}}, 
                   label=lst-error]
Entity: line 23: element capacity: Relax-NG validity error : Type double doesn't allow value 'magic'
Entity: line 23: element capacity: Relax-NG validity error : Error validating datatype double
Entity: line 23: element capacity: Relax-NG validity error : Element power failed to validate content
 ERROR(core  ):Document failed schema validation
\end{lstlisting}

\section{Conclusions}
\label{sec-conc}

Writing archetypes can be a daunting task because reasonably accurate models 
require knowledge of physics, economics, and computer science to solve a single nuclear engineering 
problem.  Unlike other spheres of nuclear engineering, decoupling these domains from
one another is often not possible without significant simplification. \cyclus is 
no exception to this and is designed to allow for complete fidelity throughout 
all aspects of the simulation. The advantage of the \cyclus design is that it 
enables full modeling fidelity without requiring that archetype developers actively 
address every class of problem every time they pursue a new archetype.

\Cyclus succeeds in simplifying archetype development by identifying a category 
of computer science and software development problems that are addressed 
algorithmically. This moves effort away from humans, who are pursing physics and
economics, and onto computers. This automation happens by default for
archetype development and reduces manual code writing by approximately
10 times.
Additionally, the automation may be partially or fully reduced 
if the default generated code is not desired. 

The \cyclus system enables better fuel cycle simulations by creating better 
archetypes.  Improved archetypes are a direct consequence of two features of the preprocessor. First, 
\cyclus encourages developers to write the archetype
that they intended to implement. Secondly, the archetypes are automatically
validated.

State variables are easy to create. When a software feature has a high cost to use,
developers will minimize the number of times that they invoke it. This can 
sometimes lead to sacrificing model fidelity in an effort to author more concise
code. However, the long-term completeness of an archetype
with respect to its physics calculations should not be based, even in part, on the
short-term impetus to have a minimum-viable product. By dramatically reducing the 
length of time it takes to implement a state variable, archetype developers implement
more state variables and thus more precise and tunable models.

Furthermore, automatically generating archetype code removes typographic errors and 
\cyclus interface misuse. This avoids potential and frustrating problems in 
archetype development. The 
generated code derives its validity from \cycpp, which itself is extensively 
tested. Any errors accidentally introduced by \cycpp would be endemic to all archetypes, 
but a fix to \cycpp would be the corresponding solution. Archetypes are thus 
partially vetted due to \cycpp.

The preprocessor also generates schema for archetypes. This provides a mechanism 
for automatically validating user input.  Without validation, the archetype 
developer has no guarantee that the user has entered a meaningful or physically possible 
value (e.g., negative fluxes). Rather than approaching this problem in an
\emph{ad hoc} 
manner, the \cyclus interface demands that user input be examined via \gls{RNG}.  The overhead 
of this requirement is mitigated since the archetype developer obtains the schema
for free. This assures a high level of quality in using archetypes as well as 
developing them.

The strategies detailed and implemented in this paper radically
reduce the overhead of writing archetypes. By enabling more expressibility and greater
modifiability, developers and simulators are able to more easily experiment.  
Alternative fuel cycle representations may be explored quantitatively at an
unparalleled rate.

However, archetype development tools and approaches are not without further 
potential refinements. \cycpp will undergo continued improvement 
as more archetypes are developed and common usage patterns emerge. Codifying and 
auto-generating these patterns is a rich area of exploration. The authors anticipate 
that inventory and resource exchange patterns will be among the first
targeted, 
which would likely take the form of new filters in pass three.

Furthermore, the fundamentals of the \cyclus type system will be used in 
perpetuity, as enabled by the extensibility of the type system.  More types
will be added as needed by the archetype developers.  It is also possible to 
make the type system dynamic, allowing for custom types to be implemented at run time.
This could be a boon to archetype developers seeking to create a wide variety of custom
tables.

The preprocessor could also improve the generated code. \Cyclus reserves the right
to add additional metadata keys and associated meanings.  For example, a \code{'range'}
key could specify the acceptable range for a state variable. This could in turn 
provide better validation by adding bounds checks beyond the what is performed
by \gls{RNG}. A similar strategy could be followed for categorical variables whereby
set membership would be verified. The code that is generated for future keys 
depends largely on the meaning ascribed to those keys. There is no limit 
to the richness of available metadata.

In summary, archetype development is has been made significantly easier. This is due in large
part to the advent of the \cyclus preprocessor. While \cycpp itself is the fruit of
a large computer science and software development effort, it is used here 
primarily to improve fuel cycle simulations. \cyclus provides a solid ground 
as a platform for archetype development while simultaneously nimble enough
to allow for future growth. Independent of \cyclus, the strategies and methods that
\cyclus implements for archetype developers are translatable to any agent-based 
fuel cycle simulator.  Some aspects, such as the type system,  may even be exportable 
to general simulation science.

\bibliographystyle{elsarticle-num}
\bibliography{refs}
\end{document}